\documentclass[a4paper,fleqn,usenatbib]{mnras}

\usepackage{newtxtext,newtxmath}

\usepackage{graphicx}	
\usepackage{amsmath}	
\usepackage{amssymb}	

\title[The \textsc{ARTEMIS} simulations]{The \textsc{ARTEMIS} simulations: stellar haloes of Milky Way-mass galaxies}

\author[A. S. Font et al.]{
  Andreea S. Font,$^{1}$\thanks{E-mail: A.S.Font@ljmu.ac.uk}
  Ian G. McCarthy,$^{1}$
  Robert Poole-Mckenzie,$^1$
  Sam G. Stafford,$^1$\newauthor
  Shaun T. Brown,$^1$
  Joop Schaye,$^2$
  Robert A. Crain,$^1$
  Tom Theuns,$^3$
  Matthieu Schaller$^2$
\\
$^{1}$Astrophysics Research Institute, Liverpool John Moores University, 146 Brownlow Hill, Liverpool L53RF, UK\\
$^{2}$Leiden Observatory, Leiden University, P.O. Box 9513, 2300 RA, Leiden, The Netherlands\\
$^{3}$Institute for Computational Cosmology, Durham University, Durham DH1 3LE, UK
}

\date{Accepted XXX. Received YYY; in original form ZZZ}

\pubyear{2020}

\begin{document}
\label{firstpage}
\pagerange{\pageref{firstpage}--\pageref{lastpage}}
\maketitle

\begin{abstract}
We introduce the ARTEMIS simulations, a new set of 42 zoomed-in, high-resolution (baryon particle mass of $\approx 2\times10^4 \, {\rm M}_{\odot}/h$), hydrodynamical simulations of galaxies residing in haloes of Milky Way mass, simulated with the EAGLE galaxy formation code with re-calibrated stellar feedback.
In this study, we analyse the structure of stellar haloes, specifically the mass density, surface brightness, metallicity, colour and age radial profiles, finding generally very good agreement with recent observations of local galaxies.  The stellar density profiles are well fitted by broken power laws, with inner slopes of $\approx -3$, outer slopes of $\approx -4$ and break radii that are typically $\approx 20$--$40$~kpc.  The break radii generally mark the transition between in situ formation and accretion-driven formation of the halo. The metallicity, colour and age profiles show mild large-scale gradients, particularly when spherically-averaged or viewed along the major axes. Along the minor axes, however, the profiles are nearly flat, in agreement with observations.  Overall, the structural properties can be understood by two factors: that in situ stars dominate the inner regions and that they reside in a spatially-flattened distribution that is aligned with the disc.  Observations targeting both the major and minor axes of galaxies are thus required to obtain a complete picture of stellar haloes. 
\end{abstract}

\begin{keywords}
galaxies: stellar content, galaxies: structure, galaxies: haloes
\end{keywords}

\section{Introduction}

Within the framework of the $\Lambda$CDM cosmology, the host dark matter haloes of galaxies like the Milky Way are thought to assemble hierarchically, through the accretion and disruption of a multitude of smaller structures, such as dwarf galaxies \citep{white1978,searle1978}. Consequently, their stellar haloes are expected to contain tidal debris in various stages of phase-mixing.  At the same time, the chemical abundance distribution of haloes is predicted to display patterns that are related to the intrinsic properties of the stellar halo progenitors (e.g.~stellar mass) and to their time of accretion. Thus, the formation history of galaxies can potentially be decoded from the present-day structural and chemodynamical properties of the stellar halo, with the outer regions of the halo being particularly information rich due to the longer timescales for phase-mixing.

However, relying only on the information imprinted in the accreted halo limits the understanding of the formation history of galaxies to only a relatively small number of stochastic events.  Aside from the accreted stars, galaxies of course also contain stars that were born `in situ', the majority of them being located in the central regions (particularly the disc), but a non-negligible fraction is also predicted to be found in the halo.  Observational results indicate that the stellar halo of the Milky Way has a `dual nature', with a more metal-rich, flattened and slightly rotating inner component and a more metal-poor, rounder and non-rotating outer one \citep{carollo2007}.  A more recent analysis of the stellar halo using Gaia DR2 data has revealed that approximately {\it half} of the total halo population in the Solar neighborhood is likely to have been born in situ \citep{belokurov2019}.  A significant population of in situ stars in the inner halo has also been revealed by the H3 survey \citep{conroy2019}. These findings lend strong credence to the early predictions made about the in situ halo using hydrodynamical simulations \citep{zolotov2009,font2011}. 

Cosmological hydrodynamical simulations are one of the most powerful methods for modeling the stellar halo.  Stellar haloes with a dual nature (that is, with both in situ and accreted components) are produced naturally in these simulations \citep{zolotov2009,font2011,mccarthy2012a,cooper2015,tissera2014,pillepich2015,clauwens2018}. Note also that the in situ halo components obtained in such simulations are conceptually quite different from the monolithic collapse model advocated by \citealt{eggen1962}, because their growth is driven by the process of hierarchical assembly (and therefore fully consistent with it) and because this growth occurs over a significant fraction of the age of the Universe.

However, cosmological hydrodynamical simulations have their own limitations. Currently, there is no clear consensus on the details of the in situ stellar halo; for example, on its spatial distribution, the fraction it contributes to the total halo, or even the origin of its stars. In some simulations, the in situ component dominates the inner $\approx 20$~kpc \citep{font2011}, while in others, it is confined mostly within $5-10$~kpc \citep{pillepich2015}.  The proposed origins of the in situ halo component also vary and  include dynamical heating of nascent discs as a result of interactions with satellites, direct star formation in the halo (e.g.~from dense gas stripped from satellites), or from gas cooling in large filamentary structures \citep{zolotov2009,font2011,pillepich2014,cooper2015}. 
The fact that simulations make different predictions for the in situ haloes of galaxies of similar mass is likely a consequence of differences in the treatment of physical processes (in particular, star formation and stellar feedback) and perhaps also the numerical resolution.

Observations can potentially be used to constrain the properties of the in situ and accreted components and therefore, implicitly, to test models of galaxy formation.  For example, empirical radial profiles of stellar mass density (or surface brightness) and of metallicity, colour, and age can provide valuable information on the physical nature of the halo, as the mixture of accreted and in situ stars is predicted to vary strongly with galactocentric distance.  These are becoming standard observational tests, which have been applied extensively to the study of the halo of the Milky Way \citep{sesar2011,deason2011,deason2012,deason2014,xue2015}, that of M31 \citep{guhathakurta2005,courteau2011,gilbert2014,ibata2014}, and of a growing number of nearby external galaxies \citep{bakos2012,monachesi2016a,harmsen2017,rich2019}.  Such quantities can also be readily derived from cosmological hydrodynamical simulations and used to assess their realism.  While some progress has previously been made in this regard, comparisons to date have generally not taken full advantage of the richness and quality of the available observational data.

From previous observational work, it has been found that the observed stellar density (or surface brightness) profiles can be well fit by a broken power law \citep{deason2011,deason2014,harmsen2017}.  The parameters, which are composed of the inner and outer slopes and the break radius, are determined by fitting to the density profiles and can be compared with fits to simulations.  Interestingly, accretion-only models (i.e., N-body simulations that contain a collisionless stellar component) also generally predict broken power laws for these profiles \citep{johnston2008,cooper2010}. Using the \citet{bullock2005} set of N-body simulations, \citet{deason2013} have shown that a special configuration of streams with apocenters all in close range can broadly reproduce the density profile of the Milky Way.  However, another possible mechanism for generating breaks in the stellar density profiles is the changing overlap between the accreted and in situ halo components. In particular, the transition between the inner, in situ-dominated component and the outer, accreted-dominated component provides a natural scale for the break radius.  So far, most of the comparisons between hydrodynamical simulations and the data have been made under the assumption of a single power law profile or by focusing on the outer part of the halo. Consequently, the power law slopes have been found to be steep, similar to the values predicted by accreted-only models \citep{pillepich2014,pillepich2018}.  It is worth noting that some observations preferentially probe the outer accretion-dominated components of stellar haloes. For example, the GHOSTS survey (e.g.~\citealt{monachesi2016a}) has a strategy of targeting only the minor axes of galaxies (or the major axis only at very large galactocentric distances), with the implicit assumption that the minor axes probe regions of the halo which are less contaminated by the disc.

The predictions for metallicity gradients appear to depend on the nature of the model (accretion-only vs.~hydrodynamical). In the accretion-only scenario, metallicity gradients are stochastic; some haloes exhibit gradients, while others do not  \citep{font2006,johnston2008,cooper2010}.  This is likely to be a strong function of the accretion history of the galaxy, the orbital parameters of the infalling satellites, and their metal content. Typically, the [Fe/H] gradients do not exceed $0.5$~dex over the scale of the halo, except in cases where massive satellites sink into the center, generating gradients of up to $\approx 1$~dex \citep{johnston2008,cooper2010}.  In contrast, some of the previous hydrodynamical simulations predicted ubiquitous metallicity gradients which are consistently large ($\approx 1$~dex over $100$~kpc) regardless of the accretion history \citep{font2011,tissera2013,tissera2014}. However, some hydrodynamical simulations display large variations between the metallicity gradients, with differences of up to $\approx 0.8$~dex out to, for example, $\approx 60$~kpc \citep{monachesi2019}.  Within the dual nature scenario, the magnitudes and slopes of the metallicity/colour/age gradients are further dependent on the abundance and spatial extent of the in situ stars and therefore can potentially be used to constrain the models.

Given the difficulty of measuring metallicities across the full extent of the stellar haloes, current observations are less conclusive with regards to the frequency of metallicity gradients. For the Milky Way, many halo samples are affected by selection biases, which may influence the metallicity measurements (see \citealt{conroy2019}). This may explain the discrepancy between the (weak) metallicity gradient found initially \citep{xue2015,carollo2007}, and a flat distribution around [Fe/H] $\approx -1.2$ determined more recently \citep{conroy2019}.
In contrast, M31 exhibits a strong metallicity gradient \citep{gilbert2014,ibata2014}, even when measured along the minor axis of the galaxy. Some of the haloes of other external galaxies observed with the GHOSTS survey display colour gradients, however others do not \citep{monachesi2016a,harmsen2017}. This calls into question the apparent universality of the metallicity gradients predicted by previous hydrodynamical simulations (e.g.~those of \citealt{font2011}). 

A source of confusion stems from the lack of a consistent comparison between simulations and observations.  Specifically, as pointed out by \citet{monachesi2016b}, simulation predictions generally correspond to spherically-averaged profiles, whereas observations often probe the minor axis (or major axis only at very large distances) alone.  To make a more consistent comparison, \citet{monachesi2016b} have used the Auriga simulations \citep{grand2017} to show that simulated haloes tend to have weaker metallicity gradients along the minor axes than when metallicities are spherically-averaged.

Note that, while observations along the minor axis increase the chances of sampling the halo uncontaminated by the disc stars, they may also be biased towards sampling the accreted component of the halo. Given that some of the in situ halo may originate in the disc, observations along the major axes (e.g.~near the disc/halo interface) can provide crucial information about the properties of this additional component.   

In order to address the above questions, we have constructed a new suite of hydrodynamical simulations of Milky Way-analog haloes called ARTEMIS, (Assembly of high-ResoluTion Eagle-simulations of MIlky Way-type galaxieS).  These simulations are run with the same hydrodynamical simulation code used for the EAGLE project \citep{schaye2015,crain2015}, but applied here at significantly improved spatial and mass resolution, using the zoom-in technique (see below).  The resolution of ARTEMIS is similar to other very recent high-resolution, hydrodynamical simulations of the Milky Way-analog haloes \citep{sawala2016,grand2017,garrison-kimmel2018,buck2020}, making possible the study of the structural properties of stellar haloes of individual galaxies and more detailed comparisons with the observational data.  Our sample of simulated galaxies is purely selected by halo mass and is larger than previous samples simulated at this resolution, providing a more diverse ensemble of merger histories that are compatible with the emergence of a Milky Way-like galaxy disc.  

The paper is organised as follows. In Section~\ref{sec:sims} we present the new suite of simulations of Milky Way-analog haloes and discuss their global properties.  In Section~\ref{sec:mainprops} we compare the simulations to observations of local galaxies (not stellar halo specific), including the total stellar masses, star formation rates, sizes, and metallicities, in order to assess the overall realism of the simulations.  In Section~\ref{sec:haloprops} we explore the structural properties of the stellar haloes of the simulated galaxies, such as their radial profiles of stellar densities, surface brightness, metallicities, colours and ages.  We also examine a number of scaling relations linking the stellar halo to the properties of the galaxy (e.g.~total stellar mass).  We test our results against a wide range of observations, including the Milky Way and external galaxies of similar mass and determine the contributions of the in situ and the accreted components to these profiles.  Section~\ref{sec:compare_sims} includes a comparison of ARTEMIS with results derived from the previous GIMIC and EAGLE simulations in order to explore the variation in the predictions of different simulations.  Finally, in Section~\ref{sec:concl} we discuss the main results and conclude.

\section{The ARTEMIS simulations}
\label{sec:sims}

The ARTEMIS suite presently comprises $42$ `Milky Way-analog' haloes simulated at high resolution with hydrodynamics, using the version of the Gadget-3 code (last described in \citealt{springel2005}) with the hydrodynamics solver and galaxy formation (subgrid) physics developed for the EAGLE project \citep{schaye2015,crain2015}.  As described below, the parameters characterising the efficiency of (stellar) feedback have been adjusted to obtain an improved match to the observed stellar mass--halo mass relation (as inferred from abundance matching), but otherwise the subgrid physics is unchanged. Below we describe our method for generating initial conditions for the simulations, the implemented subgrid physics, and we present a discussion of feedback (re-)calibration.

\subsection{Initial conditions}

Our goal is to simulate a statistically significant sample of Milky Way-analog haloes at high resolution (with a softening length of $\approx$100 pc$/h$ for all particles and a baryon particle mass of $\sim 10^4$ {\rm M}$_\odot/h$).  This is currently extremely challenging if done in the context of full periodic box runs done at high resolution and with hydrodynamics.  We therefore employ the `zoom in' technique (e.g.~\citealt{bertschinger2001}), to simulate Milky Way-analog haloes at high resolution and with hydrodynamics, within a larger box that is simulated at comparatively lower resolution and with collisionless dynamics only.

We use the \textsc{MUSIC} code\footnote{\url{https://www-n.oca.eu/ohahn/MUSIC/}} \citep{hahn2011} to generate the initial conditions of the base periodic box from which we select the Milky Way-analog haloes to be re-simulated, as well as the zoomed initial conditions.  For the base periodic box, we simulate a volume of $25$~Mpc/$h$ on a side with $256^3$ particles. The initial conditions are generated at a redshift of $z=127$ with a transfer function computed using the CAMB\footnote{\url{https://camb.info/}} Boltzmann code \citep{lewis2000} for a flat $\Lambda$CDM WMAP9 \citep{hinshaw2013} cosmology ($\Omega_m=0.2793$, $\Omega_b=0.0463$, $h=0.70$, $\sigma_8=0.8211$, $n_s=0.972$).  The initial conditions are generated including second order Lagrangian perturbation theory (2LPT) corrections.

We run this base periodic volume down to $z=0$ using Gadget-3 with collisionless dynamics and select from the completed simulation a volume-limited sample of haloes (i.e., all haloes) whose total mass lies in the range $8\times10^{11} < {\rm M}_{200,{\rm crit}}/{\rm M}_\odot < 2\times10^{12}$, where ${\rm M}_{200,{\rm crit}}$ is the mass enclosing a mean density of 200 times the critical density at $z=0$. 
This approximately spans the range of values inferred for the Milky Way from a variety of different observations \citep{guo2010,deason2012,mcmillan2017,watkins2019}.  There are $63$ such haloes in this mass range in the periodic volume. We have constructed dark matter-only simulations for each of these haloes. These simulations are discussed in more detail in \citet{poole_mckenzie2020}. Here we present results of hydrodynamic simulations for the first $42$ of these haloes, leaving the remaining haloes to be added in a future study. 

We note here that there is a slight inconsistency in the method of selection by mass, in that our selection is done on a collisionless simulation, whereas the observational mass estimates of the Milky Way generally constrain the total mass including the potential baryon effects on the halo mass itself\footnote{Note, however, that when one employs abundance matching for external galaxies, one is typically inferring the implied mass in a collisionless simulation, rather than the total mass including baryon effects.}.  If feedback is sufficiently strong to eject large quantities of baryons (which can then allow the underlying dark matter to expand), this can result in a decrease in halo mass of a given halo in a hydrodynamic simulation with respect to its collisionless counterpart (e.g.~\citealt{sawala2013,velliscig2014}).  Consequently, the masses of our final simulated haloes (zooms with hydrodynamics) are slightly lower than the quoted range above (see Table \ref{tab:table1}), but most haloes are still fully compatible with observational estimates for the Milky Way.

Note that, given the modest size of the parent periodic box, we are simulating mainly Milky Way-mass haloes in regions at/near the mean density of the Universe.  We would not, for example, simulate such haloes in close proximity to a large galaxy cluster or a deep void, since such structures are rare and not generally present in the periodic volume (which is constructed to be of mean density).  Considerably larger volumes would be required to simulate Milky Way-mass haloes near such rare fluctuations.

To generate the zoomed ICs, we first select all particles within $2 {\rm R}_{200,{\rm crit}}$ of the selected haloes and trace them back to the initial conditions of the periodic box, at $z=127$, using their unique particle IDs.  The outer radius for particle selection was chosen to ensure that we simulate, at high resolution, a region that at least encloses the splashback radius, which marks the physical boundary of the halo out to which particles pass on first apocenter (e.g.~\citealt{diemer2014}).  We choose the center of the zoom region to correspond to the median $x$, $y$, $z$ coordinates of the particles at $z=127$ and choose the lengths of the zoom region (which is a cuboid) to encompass all of the selected particles.  Inevitably, we simulate a slightly larger region than $2 r_{200,{\rm crit}}$, as the cuboid will also include a small fraction of particles not within this radius at $z=0$.  

In MUSIC terminology, the base periodic run has a refinement level of 8 and the maximum refinement level of the zoomed region is $11$.  Thus, if the entire periodic box was simulated at the highest resolution, the run would have $2048^3$ particles.  With this level of refinement, the dark matter particle mass is $1.17\times10^5$ M$_\odot/h$ and the initial baryon particle mass is $2.23\times10^4$ M$_\odot/h$.

Following the convergence criteria discussed in \citet{power2003} (see \citealt{ludlow2019} for an update), we adopt a force resolution (Plummer-equivalent softening) of $125$~pc/$h$, which is in physical coordinates below $z=3$ and in comoving coordinates at earlier times.

The resolution adopted here is therefore comparable to that of the highest resolution simulations from other groups for this mass scale.  For example, in terms of particle mass, ARTEMIS lies between resolution levels 3 and 4 (with 3 the highest) of the Auriga simulations \citep{grand2017} (note that only 3 Auriga haloes have been simulated at the highest level 3) and levels 1 and 2 (1 being the highest) of the APOSTLE simulations \citep{sawala2016} (again only a few haloes were simulated at the highest level for APOSTLE), which also use the EAGLE code. It is also comparable in resolution to the FIRE-2 simulations of Milky Way-analog haloes \citep{garrison-kimmel2018}.

As a test of zoomed IC generation, we have compared the final halo masses (${\rm M}_{200,{\rm crit}}$) of the zooms when run with collisionless dynamics to that of the corresponding haloes in the initial base periodic run, finding agreement to typically better than $1\%$.

\begin{figure*}
\centering
\includegraphics[width=0.9\textwidth]{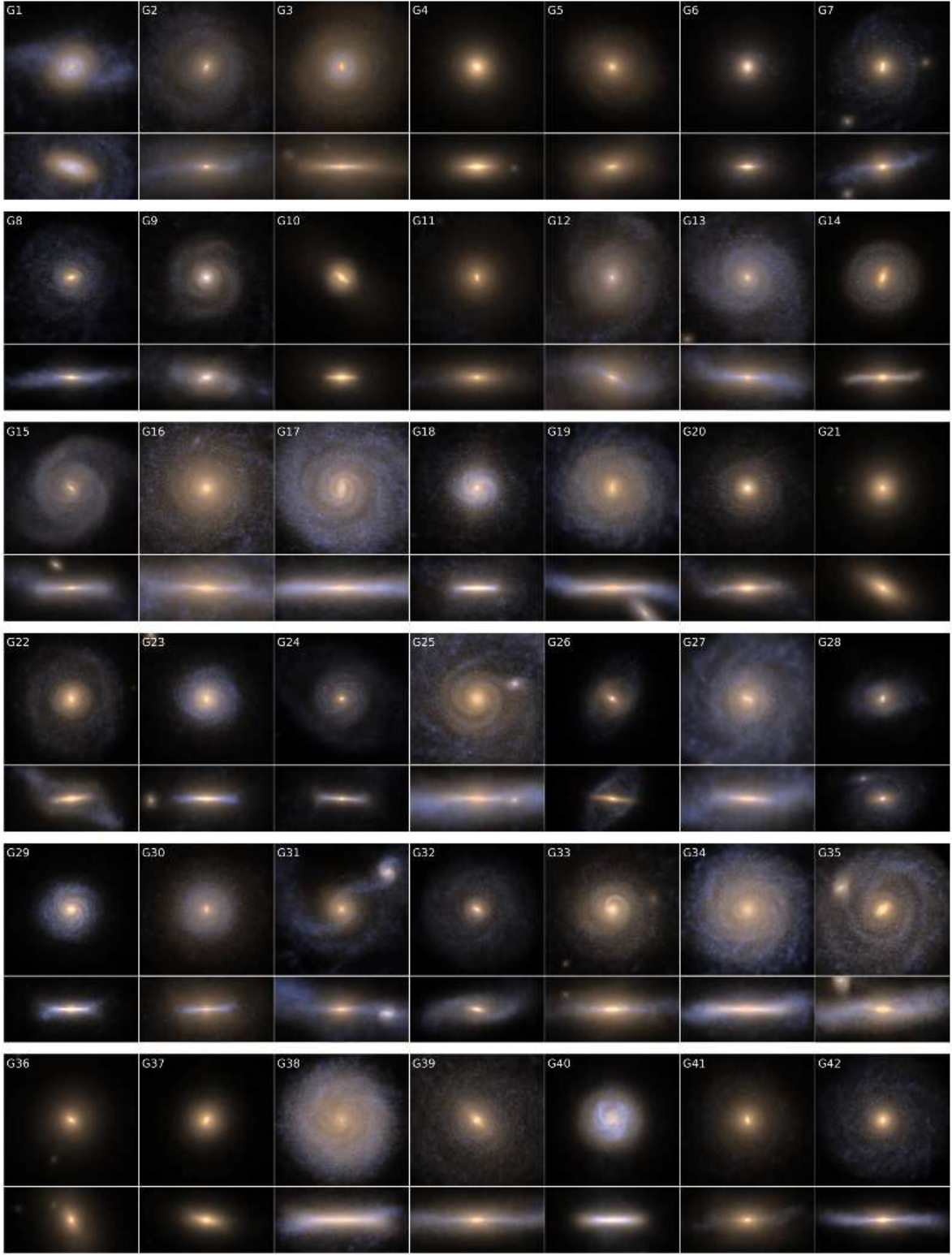}
\caption{Composite maps of the ARTEMIS systems at redshift $z=0$, showing SDSS $i$-band (red), $r$-band (green) and $g$-band (blue) luminosities. Each of the six main rows contains 2 sub rows, showing face-on (top panels) and edge-on (bottom panels) projections, respectively. Galaxies are denoted G1 $-$ G42 and are ordered from top left to bottom right. All images are $50$~kpc on the side. The maps have been constructed with the Py-SPHViewer code \citep{alejandro_benitez_llambay_2015_21703}.}
    \label{fig:sims}
\end{figure*}

\subsection{Subgrid physics and feedback (re-)calibration}

The EAGLE code includes subgrid models of important processes that cannot be resolved directly in the simulations (even at ARTEMIS resolution), including metal-dependent radiative cooling in the presence of a photo-ionizing UV background \citep{wiersma2009a}, star formation \citep{schaye2008}, stellar evolution and chemodynamics \citep{wiersma2009b}, black hole formation and growth through mergers and gas accretion \citep{springel2005b,rosas2015}, along with stellar feedback \citep{dallavecchia2012} and feedback from AGN \citep{booth2009}.  Note that at the mass scale of interest here, the AGN feedback implemented in the simulations does not play a significant role in regulating star formation.  Thus, only re-calibration of the stellar feedback is considered when attempting to match observational diagnostics.

The efficiency of the stellar feedback in the main EAGLE runs presented in \citet{schaye2015} was adjusted to approximately reproduce the local galaxy stellar mass function and the size--stellar mass relation of disc galaxies.  It was shown in that study when the resolution of the simulations was increased from the fiducial level (with a baryon particle mass of $\sim10^6$ M$_\odot$ and softening of $700$~pc) to a factor of 8(2) better mass(force) resolution, that the parameters controlling the efficiency of stellar feedback needed to be adjusted to recover a similarly good match to the calibration observables (the re-calibrated model labelled 'Recal' in \citealt{schaye2015}).  In particular, for a fixed set of parameter values, the efficiency of the feedback tends to increase with increasing resolution.  This trend of increasing efficiency with resolution is also apparent in the APOSTLE simulations \citet{sawala2016}, which used the EAGLE model but were not re-calibrated and yield stellar masses for Milky Way-analog haloes that are approximately a factor of 2 below that implied by abundance matching constraints.

Here our starting point is the recalibrated (Recal) model described in \citet{schaye2015}.  The Recal-L0025N0752 simulation (with a gas particle mass of $2.26 \times 10^{5} \, {\rm M}_{\odot}$ and a dark matter particle mass of $1.21 \times 10^6 \, {\rm M}_{\odot}$) was run in a cosmological box of $25$~comoving Mpc (on a side).  We first verified that, if we generate zooms at the resolution adopted in \citet{schaye2015}, we recover an identical stellar mass--halo mass relation to that EAGLE Recal simulation at the halo mass scale where the zooms and the periodic EAGLE volume overlap. However, as we are going to higher resolution (by approximately a factor of $7$ in mass), some re-calibration is expected to be necessary.  Indeed, through experimentation with a number of test haloes, we found that the stellar masses decreased by $\approx 0.1$ dex with respect to the EAGLE Recal-L0025N0752~box when adopting the same parameter values but run at our default zoom resolution.  Thus, a reduction in the feedback efficiency is required.  Furthermore, we note that the EAGLE Recal model itself somewhat undershoots the stellar mass--halo mass relation (i.e., predicts stellar masses that are too low at a halo mass of $\sim 10^{12} {\rm M}_\odot$) inferred from abundance matching (see \citealt{schaye2015} and also Fig.~\ref{fig:props} below), so we reduce the efficiency to achieve an improved match to the observations, rather than matching the EAGLE model. For calibrating the stellar mass -- stellar halo relation, we use the results of \citet{moster2018} and \citet{behroozi2019}, but note that the exact shape of this relation is still debated and appears to depend on galaxy morphology  (e.g. \citealt{mandelbaum2006,conroy2007,posti2019}). However, these morphological differences are most prominent at higher masses than of interest here (e.g., for massive spiral galaxies with stellar masses of $1-3 \times 10^{11}$ M$_{\odot}$, see \citealt{posti2019}). 

The feedback efficiency associated with stellar feedback in \citet{schaye2015} is a smoothly varying function of density and metallicity (see eqn.~7 of that study).  At low and high densities and metallicities the function plateaus to constant values.  The metallicity dependence is physically motivated (with increased radiative lossses for higher metallicities), but the density dependence is meant to compensate for the overcooling expected above a critical density that decreases with the numerical resolution \citep{dallavecchia2012}.
The plateau values are $0.3$ and $3.0$ times the energy available from supernovae ($10^{51}$ ergs per SN, assuming a Chabrier IMF) by default, but can be adjusted.  The transition scale from low to high and how fast this transition occurs (i.e., its slope) are also represented by adjustable parameters.  Through experimentation, we have found that the simplest way (though not necessarily a unique way) to achieve a match to the amplitude of the stellar mass--halo mass relation is to increase the density transition scale.  Specifically, we find that by increasing the density transition scale by a factor of $5$ relative to that adopted in the EAGLE Recal-L025N0752~model, we can approximately match the empirically-inferred stellar masses of $\sim10^{12}$ M$_\odot$ haloes.  Increasing the density transition scale, which can be motivated by the analysis of \citet{dallavecchia2012}, has the effect of allowing more vigorous star formation to proceed up to higher densities.

As we show below, improving the match to the stellar mass--halo mass relation in this way has additional benefits, including yielding an improved match to the sizes of discs galaxies.  A negative consequence, though, is that additional metals are produced via the extra star formation, increasing an already existing tension with the data found in \citet{schaye2015}.

Fig.~\ref{fig:sims} shows composite SDSS-like surface brightness maps for the $42$ ARTEMIS systems at redshift $z=0$, labelled G1 $-$ G42. Each of the six main rows includes two further rows, showing the face-on and edge-on views of the disc galaxies, respectively. The coordinate system is oriented such that the $z$-axis is along the direction of the total angular momentum vector $\vec{L}$ of all stars contained within a radius of $30$~kpc from the center of mass of the galaxy. Some of these galaxies are in relative isolation today, while others display ongoing interactions with satellite galaxies. We note again that the initial selection of galaxies was based solely on their total mass with no conditions on the merger history (e.g.~quiescent history), as it has been shown that Milky Way-analog disc galaxies can form via a diverse range of pathways, including also recent massive mergers \citep{font2017}. As will be shown quantitatively later, most of our simulated galaxies are disc-like. The edge-on views show also that stellar haloes have a rich inner structure, displaying tidal streams, shells and various other merger signatures that, in some cases, can extend out to at least $100$~kpc from the center. 

\section{Simulated galaxy properties}
\label{sec:mainprops}

In this section we discuss the main physical properties of the simulated galaxies and compare them with observations (e.g.~SDSS).  While the main physical properties (e.g.~total stellar mass, size, star formation rate, kinematics/morphology, etc.) are not strongly influenced by the stellar halo, it is nevertheless important to test the overall realism of the simulations.  And while the stellar halo does not significantly affect these properties, the reverse is clearly not true.  For example, if the stellar mass and size of the central main galaxy are unrealistic (e.g.~too dense and compact), one would reasonably expect the in situ component of the stellar halo to be adversely affected. 

\subsection{Main physical parameters}

Table~\ref{tab:table1} includes a number of global physical parameters of the simulated Milky Way-analog haloes, such as: total masses, total stellar masses (within $<30$~kpc), disc masses, maximum circular velocities (${\rm v}_{\rm max}$), galaxy sizes (as measured by the half-mass radius, $r_{\rm half}$), as well as the stellar co-rotation parameter ($\kappa_{\rm co}$), average stellar metallicities (both $Z_{\rm star}$ and [Fe/H]) and $V$-band magnitudes. All parameters are calculated at $z=0$.  

Star particles are separated into a disc component and the remaining (bulge + halo) component based on a kinematic criterion. Specifically, we assign them to the disc if most of their energy is associated with rotation, $K_{\rm rot,i}/{K_i} \ge 0.8$ (see Appendix \ref{sec:appendixB} for discussion; see also \citealt{font2011}). For the global rotation of each galaxy, we follow \citet{sales2010} and calculate the rotation parameter,

\begin{equation}
\kappa_{\rm rot} = \frac{K_{\rm rot}}{K}=\frac{1}{K} \sum_{i}^{r<30 \,{\rm kpc}} \frac{1}{2} m_{i} \, \big( \frac{L_{z,i}}{m_i R_i} \big)^2,
\end{equation}

\noindent where $K$ is the total kinetic energy of star particles within a radius of $30$~kpc, $K_{\rm rot}$ is the total energy in ordered rotation, $m_i$ is the mass of the star particle $i$, $L_{z,i}$ is the angular momentum of the particle along the $z$ direction and $R_i$ is the corresponding projected distance in the $x$-$y$ plane.  In Table~\ref{tab:table1} we list the co-rotation parameter, $\kappa_{\rm co},$  which is the equivalent of the $\kappa_{\rm rot}$ parameter but only summing over particles that rotate in the same direction as the total direction of rotation of the inner ($< 30$~kpc) stellar component \citep{correa2017}. The majority of our simulated galaxies have significant disc components, with $\kappa_{\rm co}>0.3$.  Note that the relation between kinematic and spatial morphology is strongly correlated (e.g.~\citealt{thob2019} and references therein) but is not one-to-one.  \citet{correa2017} found that, typically, galaxies with $\kappa_{\rm co}>0.4$ had a strong disky appearance in the fiducial EAGLE simulations.

The virial masses of the zoomed galaxies range between $\simeq 7 \times 10^{11}$ and $1.7 \times 10^{12} \, {\rm M}_{\odot},$ which are similar values to the masses of the original systems chosen to re-simulate\footnote{A notable exception is galaxy G36, which has a final total mass of $\approx 3.6 \times 10^{12} \, {\rm M}_{\odot}$. This system underwent a recent massive merger which resulted in the formation of an elliptical galaxy ($\kappa_{\rm co}=0.17$).}.  The corresponding ${\rm v}_{\rm max}$ values range between $\approx 155 - 230$~km/s.  Note that while this range of values includes the nominal rotation velocity of the Milky Way ($\approx 220$ km/s), the mean value of the simulated sample, $\approx$190~km/s, is somewhat below that of the Milky Way.  

\begin{table*}
	\centering
	\caption{The main properties of Milky Way-analog haloes in the ARTEMIS simulations. The columns include: the ID name of the simulated galaxy, the virial mass (${\rm M}_{200}$), the maximum circular velocity ($v_{\rm max}$), the fraction of stellar kinetic energy spent in co-rotation ($\kappa_{\rm co}$), the stellar mass calculated within an aperture of $30$~kpc, (${\rm M}_{\rm star}$), the mass of the stellar disc (${\rm M}_{\rm disc}$), the half-mass radius of the stellar component ($r_{\rm half}$), the median stellar particle metallicity within 30 kpc ($Z_{\rm star}$), the mean metallicity $<$[Fe/H]$>$ of the stellar component within $30$~kpc, and the $V$-band magnitude of each galaxy.}
	\label{tab:table1}
	\begin{tabular}{crccccrccc} 
		\hline
		Galaxy & ${\rm M}_{200}$ &  ${\rm v}_{\rm max}$ & $\kappa_{\rm co}$ &${\rm M}_{\rm star}$ &${\rm M}_{\rm disc}$ & $r_{\rm half}$ & ${\rm log}(Z_{\rm star}/Z_{\odot})$ & $<$[Fe/H]$>$ & $M_{V}$\\
                & $(10^{11} \,{\rm M}_{\odot})$ &  (km/s) &   &$(10^{10} \,{\rm M}_{\odot})$&  $(10^{10} \,{\rm M}_{\odot})$ & (kpc) & & & \\
		\hline
                G1  & 11.90 & 199.44 & 0.62 & 3.64 & 1.82 & 5.33 & 0.16 & -0.16 & -21.72\\
                G2  & 16.53 & 189.85 & 0.31 & 3.87 & 0.68 & 9.34 & 0.13 & -0.21 & -21.96\\
                G3  & 17.01 & 204.20 & 0.44 & 3.92 & 1.48 & 5.26 & 0.17 & -0.15 & -21.32\\
                G4  & 14.32 & 181.22 & 0.24 & 3.32 & 0.64 & 4.77 & 0.10 & -0.29 & -20.83\\
                G5  & 16.42 & 186.22 & 0.26 & 3.25 & 0.54 & 3.62 & 0.21 & -0.11 & -21.19\\
                G6  & 16.44 & 230.25 & 0.26 & 5.44 & 1.07 & 3.12 & 0.20 & -0.05 & -21.83\\
                G7  & 9.97 & 177.20 & 0.18 & 2.26 & 0.35 & 2.68 & 0.14 & -0.16 & -21.00\\
                G8  & 16.31 & 185.43 & 0.34 & 2.19 & 0.66 & 2.02 & 0.12 & -0.11 & -20.96\\
                G9  & 11.09 & 187.36 & 0.45 & 3.73 & 0.97 & 4.63 & 0.15 & -0.15 & -21.86\\
                G10 & 11.53 & 188.50 & 0.29 & 2.42 & 0.50 & 2.91 & 0.03 & -0.35 & -20.30\\
                G11  & 11.69 & 179.60 & 0.29 & 4.13 & 0.93 & 7.34 & 0.16 & -0.18 & -21.25\\
                G12  & 13.24 & 175.19 & 0.27 & 3.94 & 0.61 & 7.31 & 0.16 & -0.18 & -21.71\\
                G13  & 11.69 & 153.44 & 0.31 & 2.17 & 0.50 & 5.92 & 0.14 & -0.21 & -21.19\\
                G14  & 12.15 & 225.66 & 0.35 & 3.52 & 0.88 & 2.51 & 0.10 & -0.26 & -20.81\\
                G15  & 11.22 & 170.19 & 0.60 & 3.57 & 1.71 & 7.29 & 0.16 & -0.18 & -21.86\\
                G16  & 12.69 & 175.54 & 0.50 & 2.94 & 0.88 & 11.21 & 0.07 & -0.27 & -21.60\\
                G17  & 11.69 & 198.05 & 0.70 & 3.74 & 2.16 & 8.34 & 0.13 & -0.25 & -21.73\\
                G18  & 9.68 & 183.74 & 0.58 & 2.78 & 1.23 & 4.79 & 0.16 & -0.18 & -21.29\\
                G19  & 9.62 & 176.88 & 0.68 & 2.57 & 1.48 & 5.35 & 0.17 & -0.21 & -21.00\\
                G20  & 10.58 & 184.52 & 0.38 & 3.36 & 0.79 & 5.83 & 0.15 & -0.18 & -21.35\\
                G21  & 10.11 & 160.90 & 0.13 & 1.75 & 0.17 & 3.93 & -0.01 & -0.49 & -19.85\\
                G22  & 10.08 & 178.89 & 0.49 & 2.87 & 1.02 & 4.09 & 0.11 & -0.26 & -20.80\\
                G23  & 9.95 & 196.69 & 0.56 & 2.87 & 1.24 & 3.00 & 0.21 & -0.14 & -21.12\\
                G24  & 10.29 & 185.29 & 0.53 & 3.63 & 1.63 & 4.45 & 0.17 & -0.14 & -21.68\\
                G25  & 9.12 & 171.76 & 0.65 & 2.57 & 1.30 & 8.33 & 0.10 & -0.24 & -21.44\\
                G26  & 8.96 & 195.18 & 0.54 & 3.52 & 1.28 & 3.68 & 0.18 & -0.13 & -21.35\\
		        G27  & 7.96 & 159.55 & 0.56 & 2.56 & 1.13 & 6.47 & 0.13 & -0.20 & -21.47\\
                G28  & 7.67 & 165.97 & 0.47 & 2.39 & 0.62 & 3.35 & 0.18 & -0.10 & -21.35\\
	        	G29  & 8.82 & 210.45 & 0.62 & 3.10 & 1.54 & 2.71 & 0.18 & -0.13 & -21.26\\
	        	G30  & 8.08 & 171.58 & 0.42 & 2.68 & 0.87 & 4.72 & 0.11 & -0.22 & -21.02\\
                G31  & 8.32 & 160.32 & 0.37 & 2.09 & 0.43 & 5.66 & 0.10 & -0.24 & -21.25\\
	        	G32  & 7.88 & 155.27 & 0.61 & 2.51 & 1.06 & 5.04 & 0.13 & -0.19 & -21.18\\
	        	G33  & 7.80 & 163.05 & 0.44 & 2.64 & 0.80 & 5.98 & 0.13 & -0.24 & -20.85\\
                G34  & 7.89 & 183.40 & 0.77 & 2.85 & 1.93 & 6.46 & 0.14 & -0.21 & -21.17\\
	        	G35  & 6.82 & 164.13 & 0.45 & 1.91 & 0.76 & 6.47 & 0.02 & -0.35 & -20.62\\    
	        	G36  & 36.36 & 214.47 & 0.17 & 4.49 & 0.33 & 6.01 & 0.12 & -0.26 & -21.13\\
                G37  & 6.66 & 162.61 & 0.29 & 1.76 & 0.41 & 2.73 & 0.06 & -0.37 & -19.82\\
	        	G38  & 7.14 & 175.70 & 0.82 & 2.97 & 2.09 & 8.70 & 0.16 & -0.21 & -21.25\\
	        	G39  & 7.48 & 165.57 & 0.36 & 1.88 & 0.49 & 6.50 & 0.03 & -0.37 & -20.70\\
                G40  & 7.57 & 154.89 & 0.66 & 2.02 & 1.05 & 4.71 & 0.15 & -0.21 & -21.20\\
	        	G41  & 6.89 & 161.58 & 0.43 & 1.95 & 0.41 & 4.60 & 0.05 & -0.33 & -20.14\\
                G42  & 7.18 & 174.25 & 0.60 & 2.31 & 1.04 & 3.42 & 0.10 & -0.22 & -20.73\\
                \hline
	\end{tabular}
\end{table*}

We characterise the sizes of simulated galaxies as the projected radius that encloses half of the total stellar mass (i.e., the effective radius).  Typically, $r_{\rm half} \simeq 5$~kpc and is an increasing function of galaxy mass.  The simulations yield a size--stellar mass relation that is in excellent agreement with that measured by \citet{shen2003} (see also Section \ref{sec:scaling_rel}) for disc galaxies in the main SDSS sample.  Note that galaxy stellar masses are calculated within a spherical aperture of $30$~kpc, to approximately mimic the observationally-derived masses (see \citealt{schaye2015} for a discussion of the choice of aperture). These values are in the range ${\rm M}_{\rm star} \simeq 2-5 \times 10^{10}$ M$_{\odot}$, such that they lie on the the galaxy stellar mass--halo mass relation.  On average, the stellar disc masses are ${\rm M}_{\rm disc} \simeq 10^{10}$ {\rm M}$_{\odot}$, which are slightly below the estimated mass of the Milky Way's stellar disc, $\approx 4.1  \times 10^{10}$ {\rm M}$_{\odot}$ \citep{bland-hawthorn2016}.  However, we note that we are using a kinematic decomposition method which can result in a more restrictive selection of disc stars compared to a spatial decomposition. 

The global metallicities, $Z_{\rm star}$ and [Fe/H], correspond to the median particle metallicity, again calculated within a spherical aperture of $30$~kpc. As mentioned before, these values are higher than the average metallicities of observed Milky Way-analog galaxies.  We will discuss these parameters in more detail below.

\begin{figure*}
  \includegraphics[width=\columnwidth]{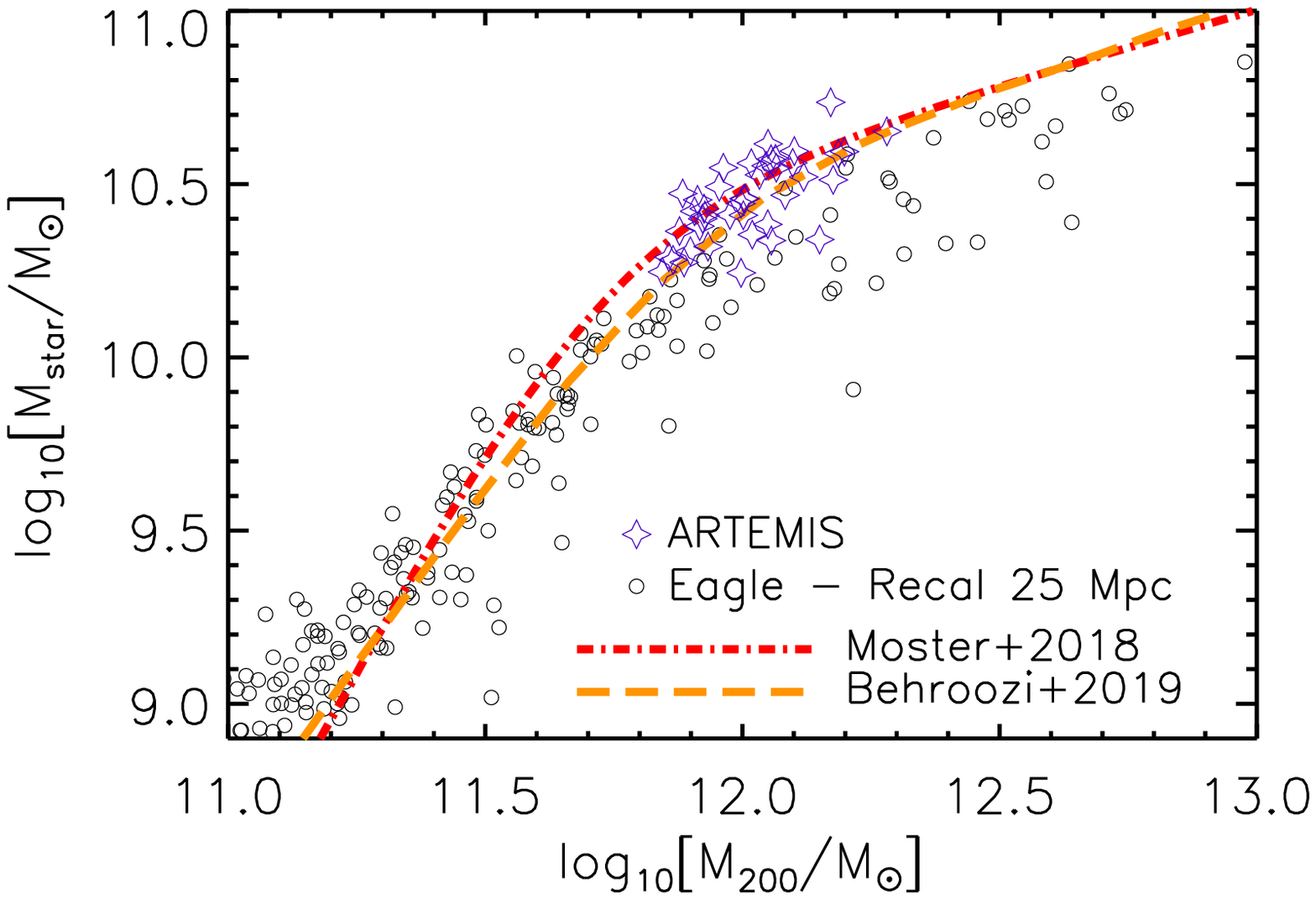}
  \includegraphics[width=\columnwidth]{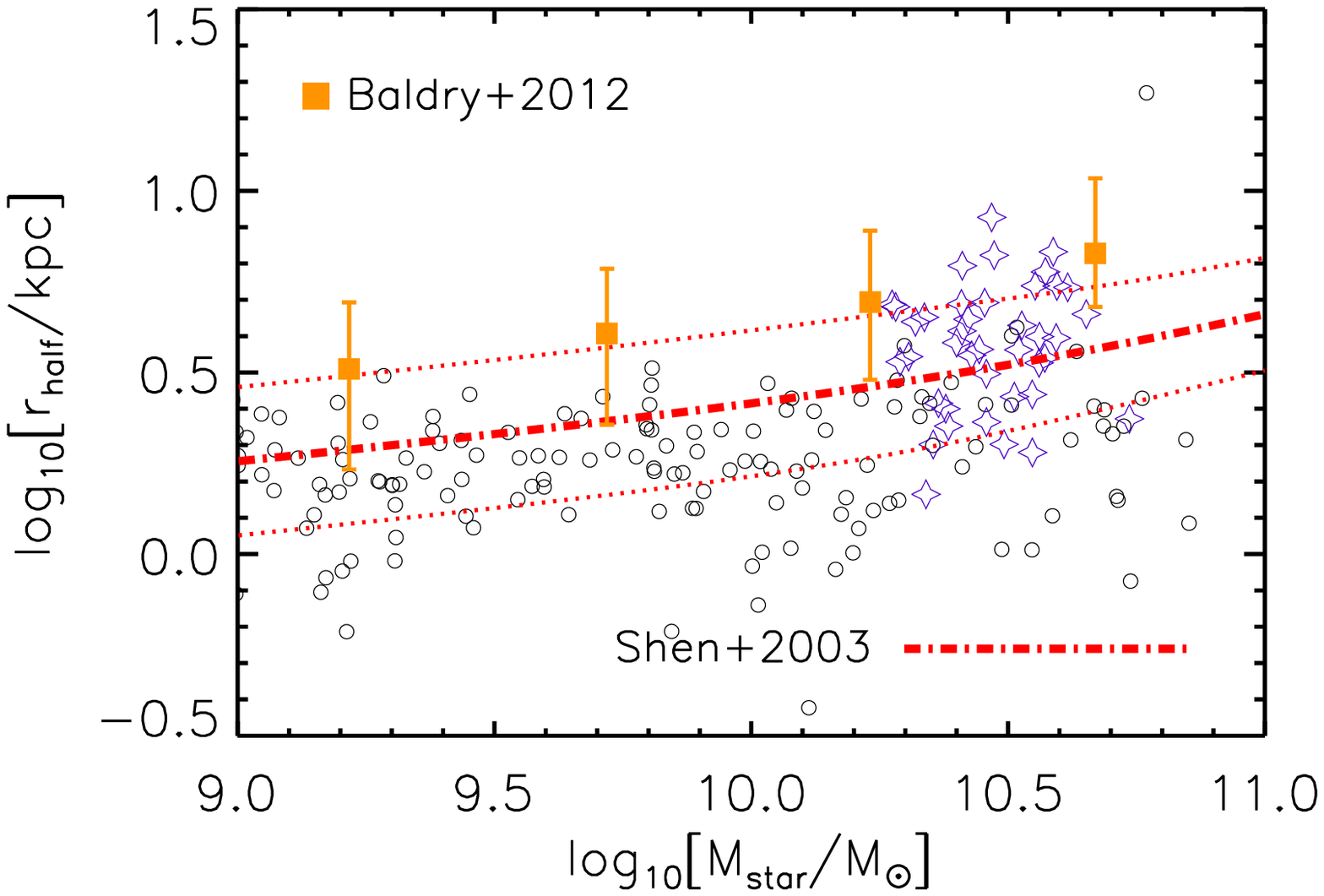}\\    
  \includegraphics[width=\columnwidth]{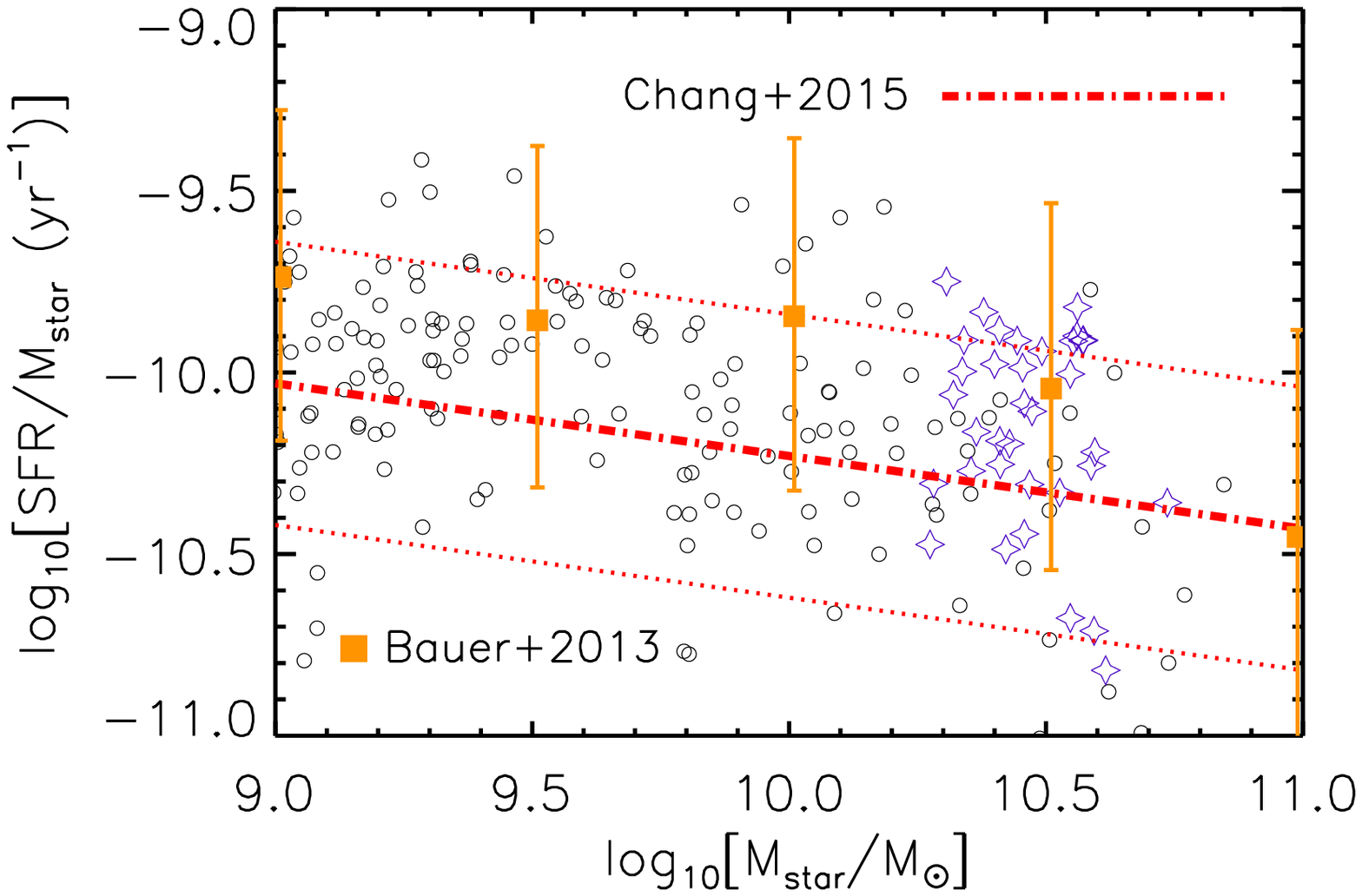}
  \includegraphics[width=\columnwidth]{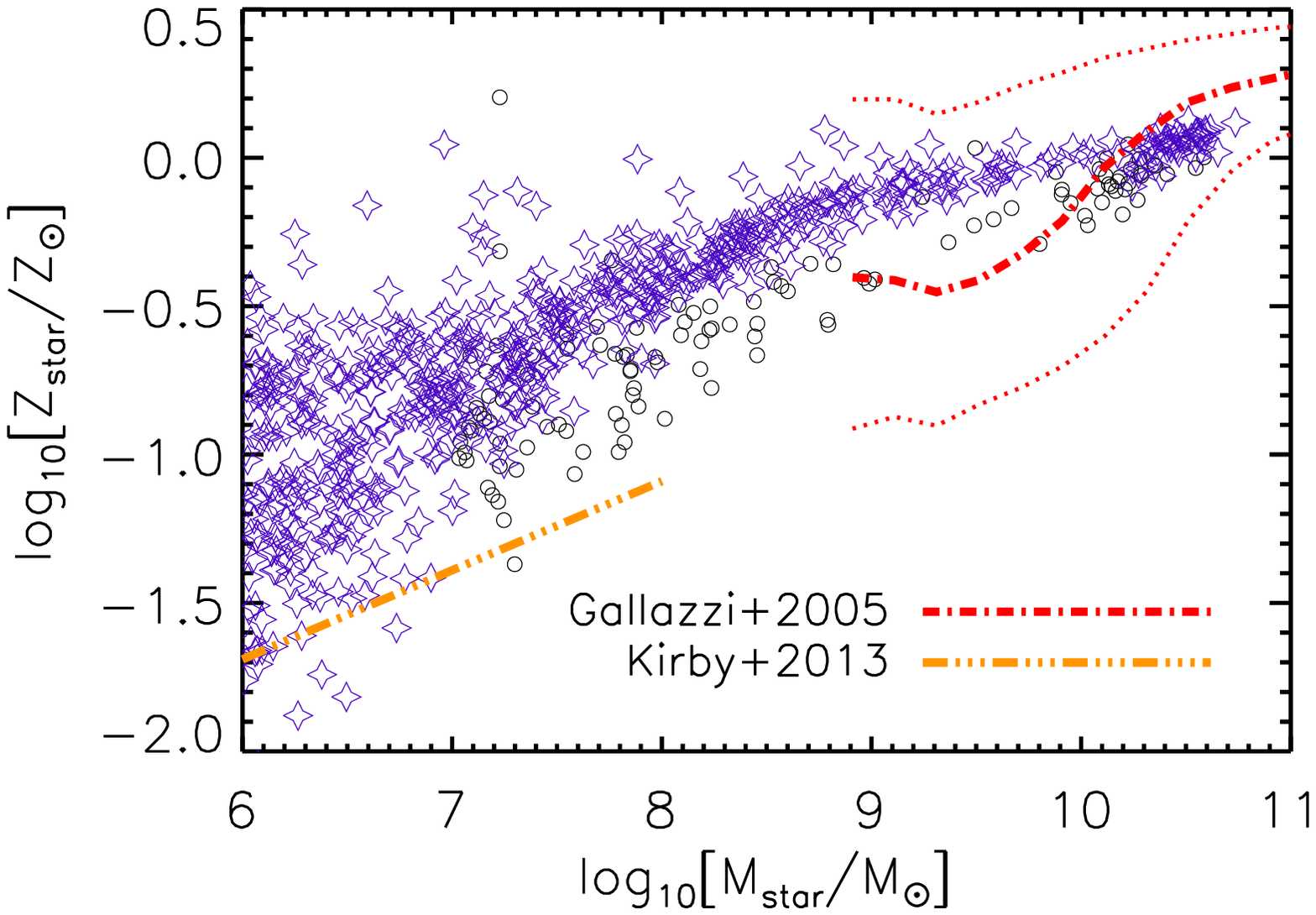}
  \caption{Comparisons of various global scaling relations between simulations and observations at redshift $z=0$: ${\rm M}_{\rm star}$--${\rm M}_{200}$ ({\it top left}), $r_{\rm half}$--${\rm M}_{\rm star}$ ({\it top right}), ${\rm sSFR}$--${\rm M}_{\rm star}$ 
  ({\it bottom left}) and the ${\rm M}_{\rm star}$--${\rm Z}_{\rm star}$ relation ({\it bottom right}), respectively. Note that ${\rm M}_{\rm star}$ is the total stellar mass of the galaxy within a spherical aperture of $30$~kpc (physical).  Galaxies in ARTEMIS are shown with purple stars and the Milky Way-mass galaxies in the EAGLE Recal~L025N0752~model with black open circles.  The dotted curves or error bars indicate the 1-sigma scatter in the observed relations. For the first three relations, we plot only central galaxies in the simulations, while for the ${\rm M}_{\rm star}$--${\rm Z}_{\rm star}$ relation we include also the satellite galaxies.  For the $r_{\rm half}$--${\rm M}_{\rm star}$ relation, we select only star-forming galaxies from the simulations, for comparison with observed relations derived for blue/star-forming galaxies.}
    \label{fig:props}
\end{figure*}

We also compute the optical light properties of the simulated galaxies in various bands. To compute luminosities, magnitudes, and colours for star particles (or for entire galaxies), we use simple stellar populations (SSPs) constructed using the PARSEC v1.2S+COLIBRI PR16 isochrones\footnote{\url{http://stev.oapd.inaf.it/cgi-bin/cmd_3.0}} \citep{bressan2012,marigo2017} and adopting the \citet{chabrier2003} stellar initial mass function (IMF) used in the simulations.  To do so, we construct a dense table of magnitudes (quoted per solar mass of stars formed) in a variety of bands (e.g.~SDSS) as a function of the age and metallicity of the SSP.  Magnitudes are interpolated for each star particle, using its age and metallicity.  The magnitudes are then converted to solar luminosities and scaled up by the total initial (i.e.~zero age main sequence) mass of the star particle.  To compute the luminosity of a galaxy, or a radial bin of a galaxy, we simply sum the luminosities of all star particles in the galaxy (or radial bin).  Note that the computed luminosities, magnitudes, and colours neglect the effects of dust attenuation. The $V$-band magnitudes, for example, of the simulated galaxies range between $-19.82$ and $-21.96.$

\subsection{Comparison with observations}
\label{sec:scaling_rel}

In Fig.~\ref{fig:props} we examine the realism of the ARTEMIS simulations against a series of observed local galaxy scaling relations; specifically, the halo mass--stellar mass, the galaxy size--stellar mass, the specific star formation rates (sSFR)--stellar mass and the stellar metallicity--stellar mass relations.  The curves correspond to various functional fits based on observational data, including the galaxy--halo fits from the abundance matching techniques\footnote{The empirical models of \citet{moster2018} and \cite{behroozi2019} predict the stellar mass at a given virial mass.  To convert the virial mass into $M_{200}$ for comparison with the simulations, we assume an NFW profile and the mass--concentration relation of \citet{dutton2014} and apply the technique of \citet{hu2003} for converting between masses of different overdensity.} of \citet{moster2018} and \cite{behroozi2019}, the effective radius--stellar mass fit for SDSS disc galaxies from \citet{shen2003} and for blue galaxies in GAMA from \citet{baldry2012}, and the sSFR--stellar mass fits of \citet{bauer2013} and \citet{chang2015} for star-forming galaxies in GAMA and SDSS+WISE, respectively.  For the metallicity--stellar mass relation above ${\rm M}_{\rm star} > 10^9 \, {\rm M}_{\odot}$ we use the fit from \citet{gallazzi2005}, complementing it at lower stellar masses with the fit obtained by \citet{kirby2013} for Local Group dwarf galaxies.

We also compare the zoom simulations (open purple stars) with galaxies in the same Milky Way mass range drawn from the EAGLE Recal-L025N0752 simulation (open black circles; particle data from \citealt{mcalpine2016}).  For a given simulated galaxy, we compute the metallicity as the median metallicity of all star particles within the central 30 physical kpc.  In Appendix \ref{sec:appendixA}, we compare this definition of metallicity with the mass-weighted mean metallicity, finding the difference between the two to be fairly large ($\approx 0.25$~dex) and dependent on stellar mass.  Which definition should be adopted is debatable.  For example, for comparison to integrated light measurement, the mass-weighted mean might be more appropriate.  For comparison to estimates based on a survey of individual stars, the median star particle metallicity might be more appropriate.

Note that even though the same code was used for ARTEMIS and the EAGLE Recal simulation, the resulting scaling relations differ.  This is a result of re-calibration of the stellar feedback in ARTEMIS and perhaps also its increased numerical resolution. 

Since the ARTEMIS simulations are calibrated to the halo mass--stellar mass relation, this relation (specifically its amplitude) is matched by construction,  at a stellar (halo) mass scale of  $\sim 10^{10.5}\, (10^{12})\, {\rm M}_{\odot}$ (top left panel). Re-calibrating to match the stellar masses also has the positive effect of yielding an improved match to the galaxy sizes, without explicit (re-)calibration to obtain this result (see top right panel of Fig.~\ref{fig:props}).  The specific star formation rates, ${\rm sSFR} = \log_{10} ({\rm SFR}/{\rm M}_{\rm star})$, in the zoom simulations also agree well with observations of star-forming galaxies at ${\rm M}_{\rm star} \sim 10^{10.5}\, {\rm M}_{\odot}$, both in terms of the typical value and the scatter.  Note that several simulated galaxies do not show in this plot as they are presently quenched (i.e. their specific star formation rates are below the ${\rm log_{10}}({\rm sSFR}) \simeq -11$ threshold). 

The bottom right panel in Fig.~\ref{fig:props} shows the predicted stellar mass--metallicity relation, ${\rm M}_{\rm star}$ - ${\rm Z}_{\rm star}$, and compares with the observations of \citet{gallazzi2005} and \citet{kirby2013}.  Since we are interested (later) in isolating the accreted component from the in situ component of the stellar haloes, we include also the low-mass satellite galaxies of the Milky Way-analog systems for comparison (for both the ARTEMIS simulation and the EAGLE Recal model). Both sets of simulations show a discrepancy with the observations, which is mostly apparent below ${\rm M}_{\rm star} \sim 10^{10}\, {\rm M}_{\odot}$, where the simulations start to diverge from the \citet{gallazzi2005} fit. The shallower slope of the simulated ${\rm M}_{\rm star}$--${\rm Z}_{\rm star}$ relation has been noted before for EAGLE \citep{schaye2015,derossi2017}, and here we see that the trend is exacerbated for ARTEMIS and continues down to lower masses. As described above, the new (re-)calibrated stellar feedback model has led to increased star formation (by construction), which has in turn led to enhanced metal-enrichment. The difference between the simulated and observed metallicities reaches up to $\ga 0.5$~dex in the classical dwarf galaxies regime. The dot-dashed line shows the fit obtained by \citet{kirby2013} for Local Group dwarfs, ${\rm log} (Z_{\rm star}) \propto 0.3 \cdot {\rm log} ({\rm M}_{\rm star})$. The intrinsic scatter in this relation (not shown) is about $0.2$~dex.

It is worth highlighting that comparison of metallicities from simulations and observations is not completely straightforward. First, there are known systematic effects in the various methods of measuring metallicities in the observations, which may result in different slopes and zero-points in the mass--metallicity relation.  This can possibly be seen in Fig.~\ref{fig:props}, where the metallicities of SDSS galaxies in \citet{gallazzi2005} have been inferred using Lick indices while \citet{kirby2013} used spectroscopic measurements.  Furthermore, the former use integrated light measurements, while the latter examine a number of resolved stars in each system.  The two derived relations do not appear to smoothly track the same underlying relation, although the lack of overlap in stellar mass of the two samples does not make this statement conclusive.  Also, the role of environment may be a factor.  Note that the SDSS observations include galaxies in diverse environments, whereas the dwarf galaxy data is mostly derived from Local Group systems.

There are also systematic effects in the simulations.  Aside from the issue of median vs.~mass-weighted mean metallicities discussed above, the nucleosynthetic yields and Type Ia supernovae rates used in the EAGLE code are uncertain by at least a factor of two \citep{wiersma2009b}, which means that there is some freedom to shift the predicted metallicities downwards (or upwards).  Changing the slope of the simulated mass--metallicity relation would likely require altering the feedback model, such that the feedback is more efficient at preferentially ejecting metals from lower-mass galaxies.  It is possible, for example, to make the metal mass loading associated with stellar feedback to be a separate independent parameter from the overall wind mass loading (as done in Illustris, \citealt{vogelsberger2013}), which can be motivated on physical grounds (e.g., \citealt{maclow1999}).

It is worth noting that these issues are not unique to simulations based on the EAGLE code. For example, the IllustrisTNG simulations also obtain more metal-rich galaxies than observed \citep{nelson2018}.  Interestingly, these authors suggest that the disparity seen in the mass--metallicity relation can be accounted for by considering the effects introduced by the different methods used for deriving metallicities in the simulations and the observations. Specifically, using dust radiative transfer calculations to compute the emergent spectrum from their simulated galaxies and then applying the spectral line (such as $D4000n$, H$\beta$, [Mg$_2$Fe]) analysis of \citet{gallazzi2005} to the synthetic spectra brings the simulations into significantly better agreement with the SDSS observations (see Fig.~2 of \citealt{nelson2018}). 

We defer the analysis of all these systematic effects in ARTEMIS to a future study.  In Section \ref{sec:compare_sims} we discuss the relation between the steepness of the mass--metalliciy relation and the metallicity gradients of stellar haloes.

 \begin{figure*}
    \includegraphics[width=\columnwidth]{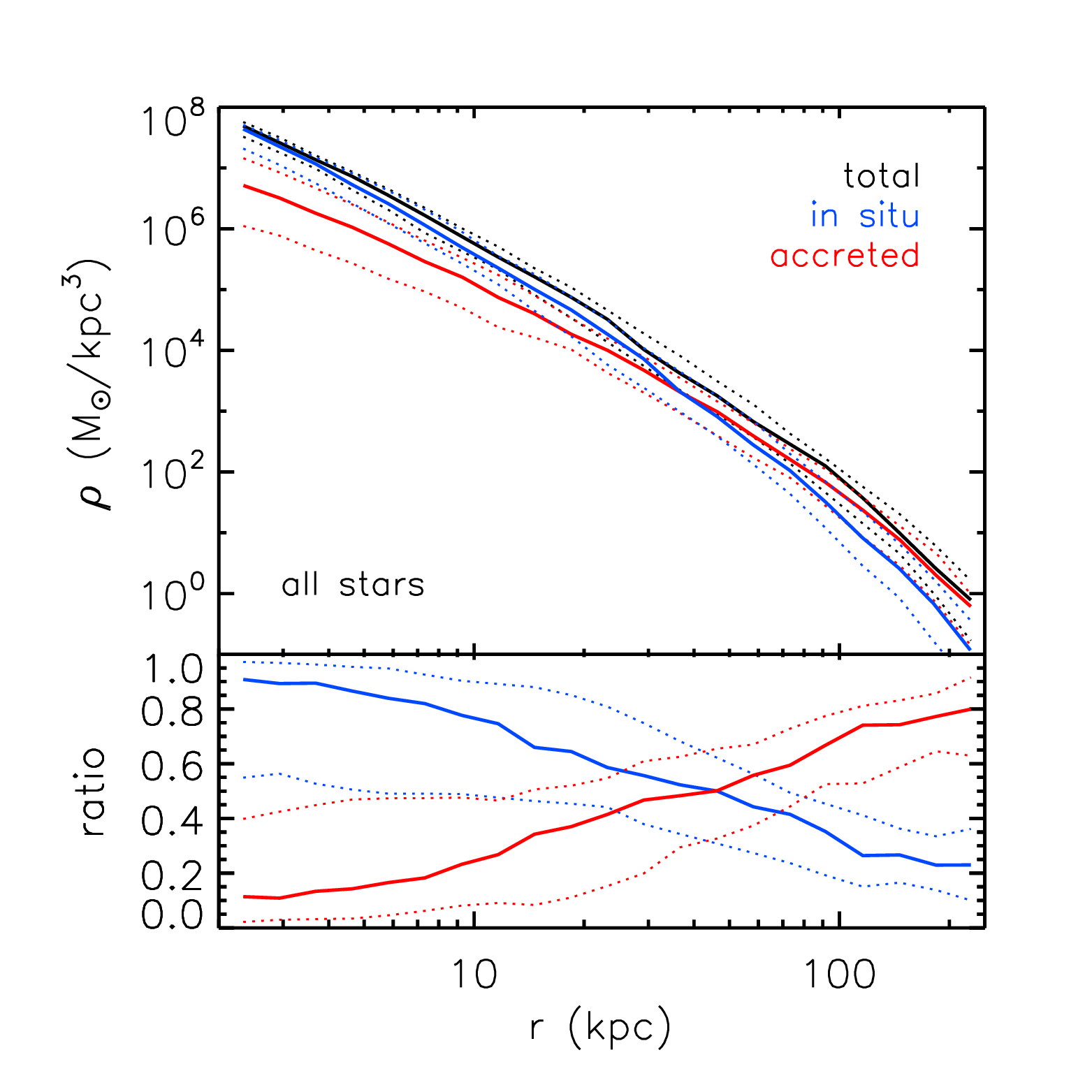}
    \includegraphics[width=\columnwidth]{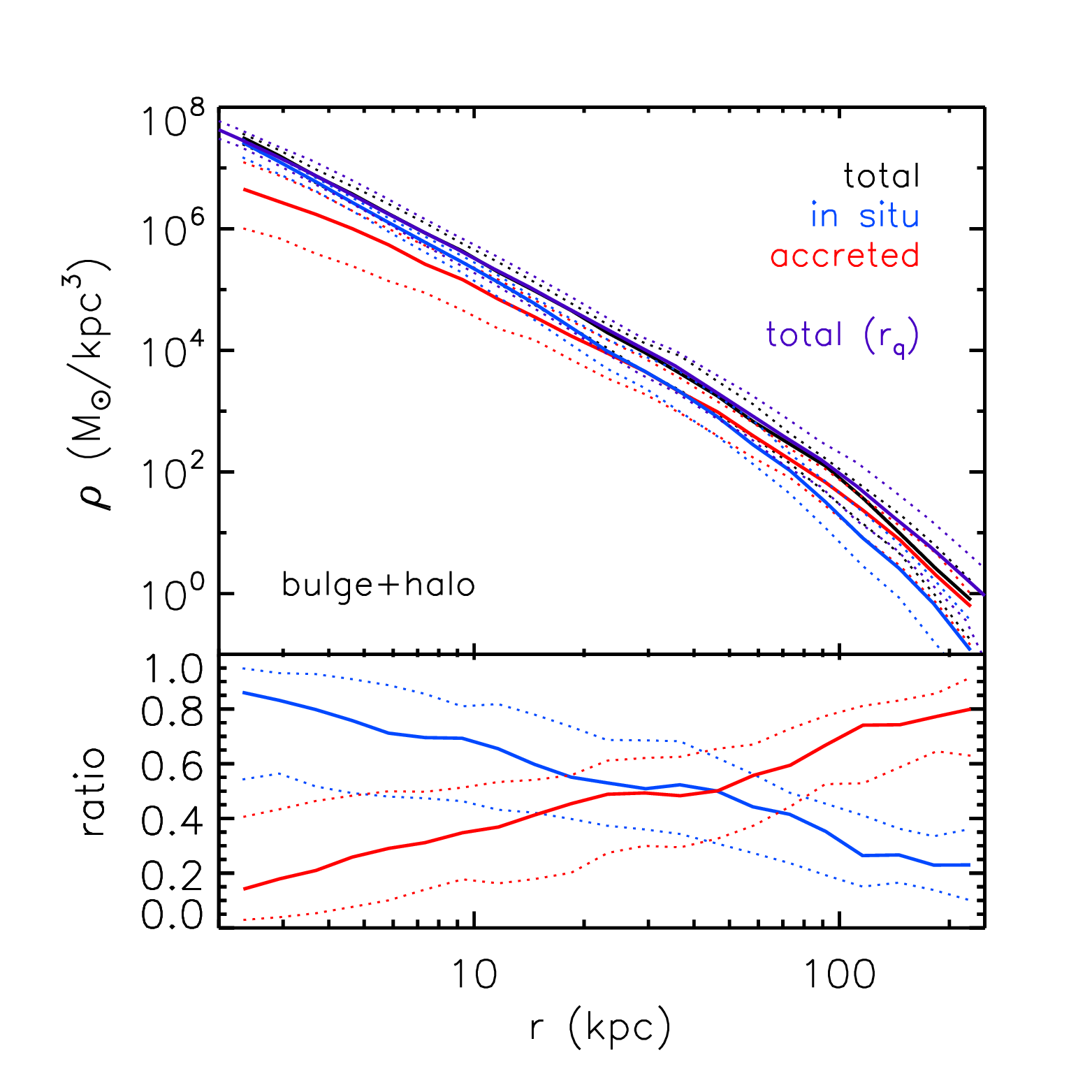}
        \caption{{\it Left:}  The median, spherically-averaged, stellar density profiles of the ARTEMIS systems for the total stellar component (black curve), in situ stellar component (light blue curve) and accreted stellar component (red curve). The dotted curves correspond to the scatter ($16^{\rm th}$ and $84^{\rm th}$ percentiles). The bottom subpanel shows the median ratios of the in situ and accreted components both with respect to the total, as well as the 1-sigma halo-to-halo scatter in these profiles. 
        {\it Right:} In analogy to the left panel but for the halo + bulge component only (i.e., omitting the disc).  The purple curve corresponds to the total density profile of the halo + bulge, corrected for the flattening, $\rho(r_q)$.}
    \label{fig:densprof}
\end{figure*}
 
\section{The structure of stellar haloes}
\label{sec:haloprops}

In this section we investigate the predicted structure of simulated stellar haloes in ARTEMIS. In Section \ref{sec:radial} we study the radial profiles of stellar densities, metallicities, colours and ages of stars in the simulated galaxies and determine the contribution of stars formed in situ to the properties of these profiles. We also make a detailed comparison with observational data. In Section \ref{sec:halo_scalings} we focus on scaling relations linking the global properties of stellar haloes to properties of the host galaxy (e.g.~total stellar mass) and compare our results with observations.  Note that, because some observations target specifically the minor and major axes of galaxies (e.g.~the GHOSTS survey), we also present results for these axes independently.

\subsection{Radial profiles}
\label{sec:radial}

\subsubsection{Stellar density profiles}
\label{sec:densprof}

The left panel of Fig.~\ref{fig:densprof} shows the median density profile for all stars (black) and all stars that are either accreted (red) or formed in situ (light blue).  The median profiles are estimated by computing the density profiles of each ARTEMIS halo individually and then simply taking the median result in radial bins.  The corresponding dashed curves represent the scatter ($16^{\rm th}$ and $84^{\rm th}$ percentiles). 

The procedure for separating stars into accreted and in situ is as follows. First, we adopt a simple definition for whether a star particle was born in situ, which is if the star particle was created in the main (most massive) subhalo of the main progenitor friends-of-friends (FoF) group of the simulated halo.  At a given earlier snapshot, we identify the progenitor FoF groups by selecting all dark matter particles within ${\rm R}_{200}$ at $z=0$ and use their unique IDs to match them to the particles of each FoF group at the earlier snapshot.  We designate the main progenitor FoF group as the one containing the largest fraction of the selected $z=0$ dark matter particles.  If a star particle is born in the most massive subhalo of this main progenitor FoF group, we designate it as having formed in situ. The left panel of Fig.~\ref{fig:densprof} shows that, by mass, the in situ stars dominate the inner regions of the galaxies, while the accreted stars dominate the outer regions. 

The total and in situ stellar profiles rise sharply towards the inner regions, tracing mostly the disc. In the outer region, the total stellar density profile falls off more steeply than the dark matter (not shown). This behaviour is as expected, and it has been explained before using accretion-only stellar halo models \citep{bullock2005,cooper2010}. Our results show that stellar haloes formed in full hydrodynamical simulations behave similarly, because the in situ stars are more centrally concentrated and the outer regions of galaxies are mainly of accreted origin. 

To determine the distribution of in situ and accreted stars that are not in the disc component, we also calculate the density profiles only for the `halo + bulge' component. In the right panel of Fig.~\ref{fig:densprof} we plot the median stellar density profile of all stars in the halo + bulge (full black line) and the median profiles for the accreted and in situ stars in the same component. Dotted curves represent the corresponding scatter ($16^{\rm th}$ and $84^{\rm th}$ percentiles) in these profiles.  Even with the disc excluded, the halo + bulge component is dominated by in situ star formation in the inner regions ($r \la 30$ kpc).  The scatter in the accreted component is larger than for the in situ component, particularly in the inner region. This suggests that the inner regions are more susceptible to stochastic events (e.g.~a few massive satellite galaxies sinking into the center can steepen the density profile). 

In terms of the in situ component, we find that the fraction of in situ stars, $f_{\rm in\,situ}$ (defined as the mass of stars that are formed in situ versus the total mass of stars, in the halo + bulge component) ranges from $\approx 70\%$ in the `Solar neighborhood' (i.e. $5\, {\rm kpc}< r< 10\, {\rm kpc}$) to $\approx 50\%$ out to $r \approx 30 - 40$~kpc. In comparison, the Milky Way's `Solar neighborhood' contains $\sim 50\%$ stars with `in situ'-like properties (e.g.~metal-rich and rotating) \citep{belokurov2019}. Recent measurements from the H3 survey also indicate a significant fraction of in situ stars in the inner halo ($25\%$ for $6 <R_{\rm gal} <10$~kpc). The halo of M31 may also have a high fraction of in situ stars, as inferred from the disc-like kinematics and metal content at the disc--halo interface \citep{dorman2013}. In particular, a component of the M31 halo appears to have the form of an `extended disc' which rotates out to distances of $\approx 40$~kpc \citep{ibata2005}.  We also note that, at least on a qualitative level, the in situ contribution and its fall off with radius in \textsc{ARTEMIS} is consistent with that found in the Auriga simulations (see fig.~2 of \citealt{monachesi2019}) and in the IllustrisTNG simulations (see fig.~14 of \citealt{merritt2020}), both of which were run using the AREPO code.

The profiles presented in Fig.~\ref{fig:densprof} were computed assuming spherical symmetry, but in general we expect the haloes to be flattened (particularly in the inner regions).  Therefore we also calculate the median stellar density profiles taking into account the flattening, $q$, which is computed following the approach described in \citet{mccarthy2012a}. For this, the stellar mass distribution is assumed to be an oblate single power law distribution of the form (in cylindrical coordinates):

\begin{equation}
\rho(R,z) = \frac{\rho_0}{[R^2 + (z/q)^2]^{\gamma/2}},
\label{eqn:rho}
\end{equation}  

\noindent where the flattening is defined as $q \equiv c/a$ and $c$ and $a$ are parallel and perpendicular to the angular momentum vector ($z$-axis), respectively. We restrict our fit to the $r<30$~kpc radius since the assumption of a single power law begins to break down on scales exceeding this value. 

We find that the median flattening of the simulated stellar bulge + halo components is $\simeq 0.63$, which agrees well with the measured value for the Milky Way stellar halo of $q\approx 0.7$ \citep{sesar2011}, with the flattening of the inner halo of M31 of $q\approx 0.6$ \citep{ibata2005}, with the median value of $q\approx0.57$ at $r<25$ kpc of the GHOSTS sample \citep{harmsen2017}, and with the average flattening ($q\approx 0.6$) of stacked stellar haloes of edge-on spirals in the SDSS \citep{zibetti2004}. This agreement is remarkable, given that no aspect of the simulations has been adjusted to reproduce the observed flattening.  By contrast, stellar haloes produced by accretion alone do not in general produce smooth, flattened (oblate) distributions (see, e.g.~fig. 6 of \citealt{cooper2010}).

The purple curve in the bottom panel of Fig.~\ref{fig:densprof} shows the median $\rho(r_q)$ profile for the halo + bulge component and the associated scatter, where  $r_q = \sqrt{r^2 + (z/q)^2}$. Overall, however, there is no significant difference between the stellar density profiles corrected for flattening and the ones assuming spherical symmetry (although differences may become apparent in individual cases).

Following on from the discussion in Section 1, we now investigate whether the halo + bulge density profiles are well fitted by a spherical broken power law (BPL) and, if so, whether the respective break radii correspond to the transition from the in situ dominated, inner region to the accreted dominated outer one. We analyse the simulated galaxies individually and use a BPL of the form:
\begin{equation}
\rho(r) \propto
\begin{cases}
r^{\gamma_{\rm inner}} \quad {\rm if} \, r \leq r_b \, {\rm kpc} \\
r^{\gamma_{\rm outer}} \quad {\rm if} \, r > r_b \, {\rm kpc}, 
\end{cases}
\end{equation}

\noindent where $r$ is the 3D spherical radius, $r_b$ is the break radius, $\gamma_{\rm inner}$ is the slope for the power law in the inner region and $\gamma_{\rm outer}$ is the slope for the outer region.

As it has been shown that accreted-only haloes can also be fitted by BPLs \citep{johnston2008,cooper2010,deason2013}, we also fit a BPL to our accreted-only halo + bulge components. Our conjecture is that, if the best-fit BPL parameters ($\gamma_{\rm inner}, \, \gamma_{\rm outer}, \, r_b$) for the total density profile are very different from the corresponding best-fit parameters for the accreted-only profile, then the in situ stars play an important role in determining the final broken-power law shape. If, on the other hand, these two sets of parameters are similar, then this implies that the resulting BPL shape is mainly the result of accretion. 

\begin{figure}
\includegraphics[width=\columnwidth]{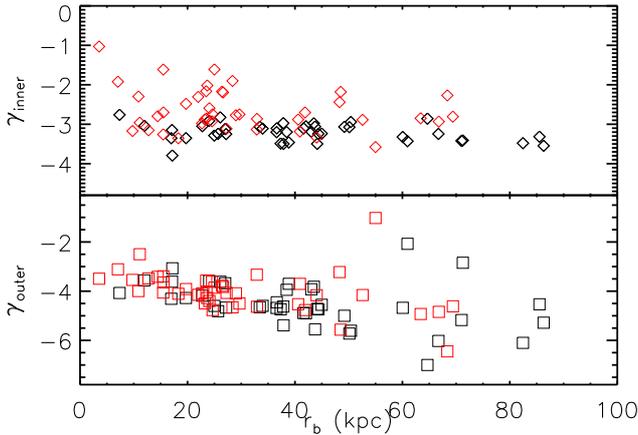}
\caption{{\it Upper:} Inner stellar mass density slopes versus break radii, $r_b$, for the best fits to individual stellar halo + bulge components. Black symbols correspond to the BPL fits to the total (in situ + accreted) stellar density profiles and red symbols to the accreted-only stellar density profiles. {\it Bottom:} The outer slopes versus break radii for the best BPL fits to individual stellar halo + bulge components.}
\label{fig:slopes}
\end{figure}
    
In Fig.~\ref{fig:slopes} we plot the best-fit BPL parameters in two panels: $\gamma_{\rm inner}$ versus $r_b$ (top) and $\gamma_{\rm outer}$ versus $r_b$ (bottom). In each panel we include results for both the total (black) and accreted-only (red) halo + bulge components. The inner slopes are found to be steeper for the total component (on average, $\gamma_{\rm inner} \approx -3.2$) compared to the accreted component ($\gamma_{\rm inner} \approx -2.6$).  Recall that the in situ fractions are high in the inner regions, which means that the inner slopes are mostly determined by the in situ component.  Observational measurements of inner slopes in galaxy haloes can therefore be used to test the predictions of hydrodynamical models; e.g.~steeper slopes ($\gamma_{\rm inner} \leq -3)$ would imply that galaxy haloes have a significant fraction of in situ stars, as predicted by the hydrodynamical simulations.  

Note that the scatter in the inner slope, $\gamma_{\rm inner}$, for the accreted component is quite large (values range between $-1$ and $-3.3$). In contrast, the scatter in the $\gamma_{\rm inner}$ of the total halo + bulge is significantly smaller, with the inner slopes values being tightly clustered around the $\approx -3.2$ value with a scatter of only a few tenths. This difference in the predicted $\gamma_{\rm inner}$ scatter can also be used to test the models.  For example, if the observed stellar haloes were built mainly through accretion, we might expect to see a large variation in the measured inner slopes among different galaxies. This would not be the case, however, if the inner haloes have a significant fraction of in situ stars. 

On average, the outer slopes are quite similar, with $\gamma_{\rm outer} \approx -4.5$ for the total halo + bulge and $ \approx -4$ for the accreted-only component.  This is expected since the outer regions are dominated by accreted stars.  However, on an individual basis, the outer slopes range between $\approx -2$ and $\approx -6$ and there is a mild anti-correlation with the break radius.
The largest scatter in the outer slope value occurs for the fits with break radii of $r_b > 40$~kpc.  One plausible reason for this large variation is the diversity of accretion histories. For example, using simulations of stellar haloes built only from accretion, \citet{deason2014} have shown that shallow outer slopes are found preferentially in galaxies that sustained more recent accretions. We note, however, that systems for which the best BPL fits return large $r_b$ values can usually also be well fitted by single power laws (SPL). Other hydrodynamical simulations, like Auriga, also include systems which are well fit by SPLs \citep{monachesi2019}.

\begin{figure*}
  \includegraphics[width=\columnwidth]{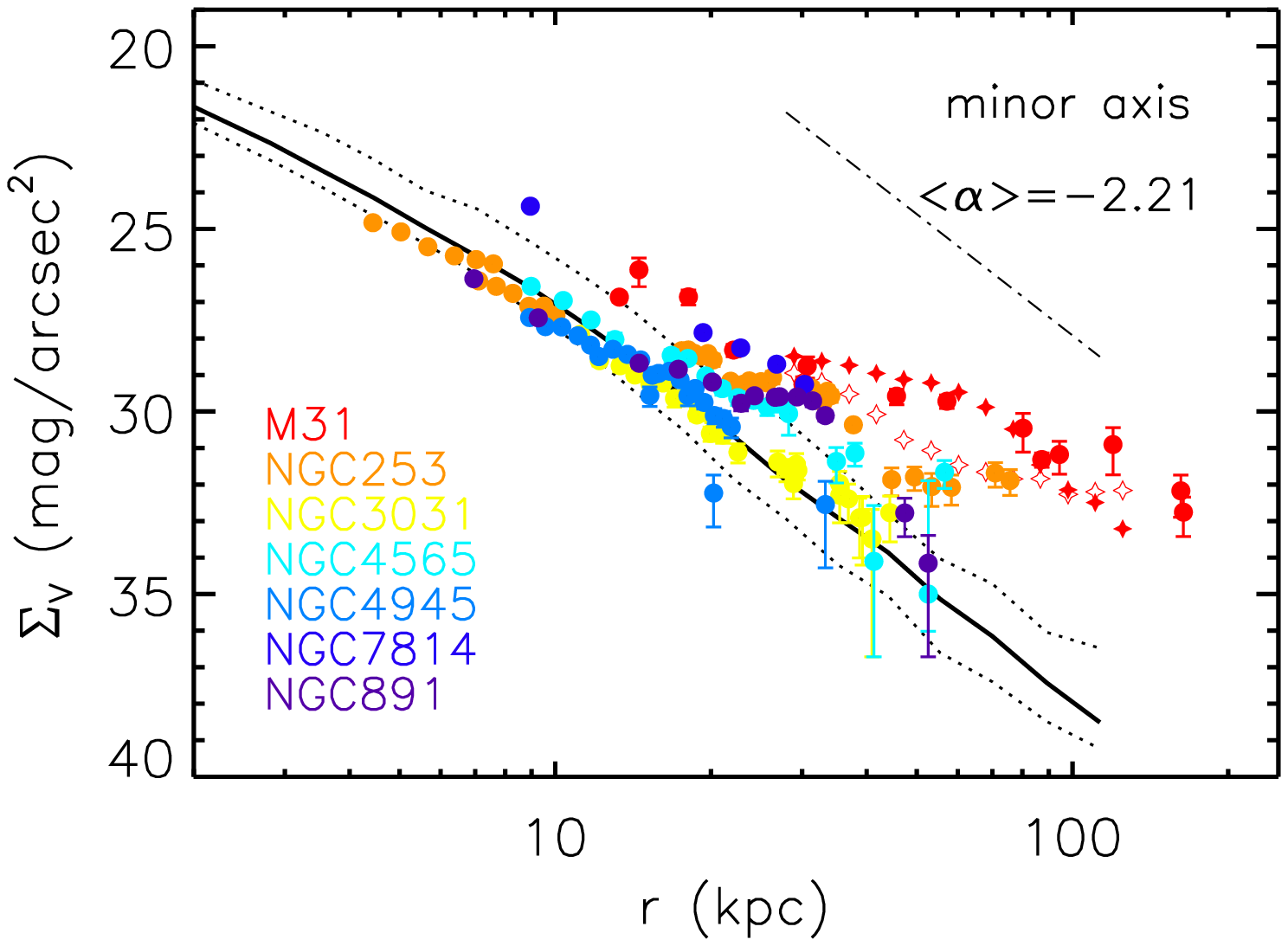}
  \includegraphics[width=\columnwidth]{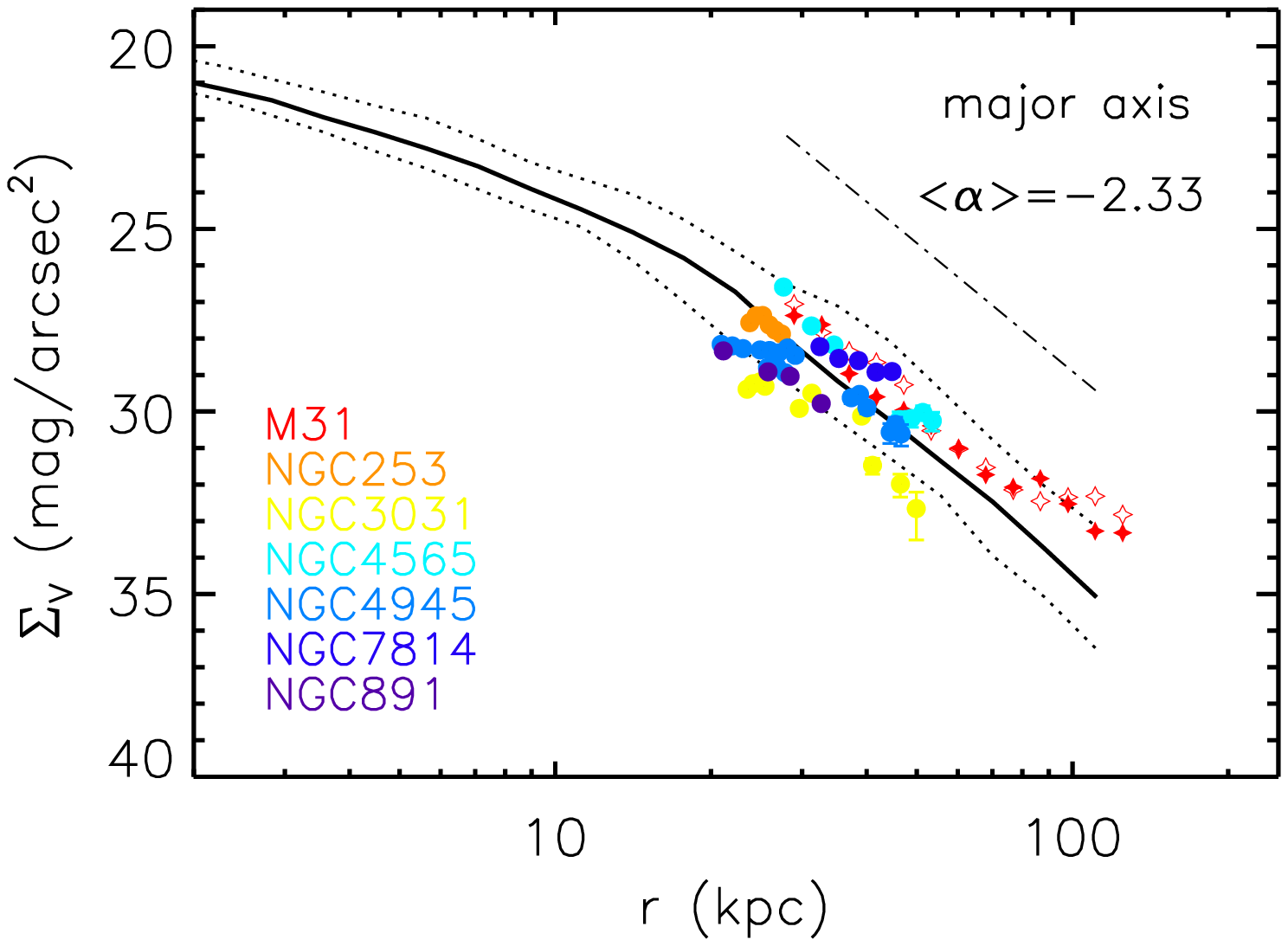}
  \caption{{\it Left:} $V$-band surface brightness profiles, $\Sigma_{V}(r)$, along the minor axes. The solid black curve shows the median profile of galaxies in ARTEMIS (computed by averaging the individual profiles) with dotted curves corresponding to $16^{\rm th}$ and $84^{\rm th}$ percentiles. The dot-dashed curves shows the best power law fit to the median profile beyond $25$~kpc which has a slope of $-2.21$. The variously coloured data points correspond to the GHOSTS disc galaxies observed along their minor axes (NGC253, NGC3031, NGC4565, NGC4945, NGC714, NGC891; data from \citealt{harmsen2017}) and M31. For M31, we plot data for the SE region, with measurements of \citet{gilbert2012} shown with filled red circles and those of \citet{ibata2014} with filled red stars, and in the NW region with measurements from \citet{ibata2014} shown with empty red stars. {\it Right:} Similar to the left panel, but for the major axes of galaxies. For M31, we plot the measurements along the major axis in the NE (empty red stars) and SW (filled red stars) regions. The best power law fit to the median simulated profile beyond $25$~kpc has a slope of $-2.33$.}
    \label{fig:surfV_data}
\end{figure*}

The best-fit break radii have a median value of $r_b \approx 41.5$~kpc for the total halo + bulge component and $\approx 30$~kpc for the accreted component, respectively.  However, the spread in the $r_b$ values is quite large. This behaviour has been found before for the accreted-only haloes, for example in the study of \citet{cooper2010} who found BPL break radii that vary between $10$ and $100$~kpc. In this case, the large variation in $r_b$ can be associated with the specifics of the accretion history.  However, our simulated haloes contain both in situ and accreted stars and, typically, the in situ component dominates out to $\approx 35$~kpc (see Fig.~\ref{fig:densprof}). In this case, $r_b$ may be associated with the transition from the in situ-dominated to the accreted-dominated region. Overall, the large variation suggests that there is not a typical radius at which this transition occurs, even for systems of similar dark matter halo mass.  It would be interesting to dissect the physical origin of the scatter in these trends by correlating the profile parameters with parameters that characterise the mass accretion histories of these systems (e.g., such as formation time, binding energy, time of last major merger, etc.).  We leave this for future work.

Observations have found that the stellar halo of the Milky Way is well fitted by a broken power law, with an inner slope of $\approx -2.5$, a steep outer slope of $-4$ to $-5$, and a break radius of $\approx 25$--$30$~kpc \citep{watkins2009,deason2011,sesar2011,xue2015}.  However, as a word of caution, the reported inner slope for the Milky Way was derived from samples that may have missed a significant fraction of the in situ component, which has only been uncovered recently using Gaia data\footnote{Most existing Milky Way-based studies of the stellar halo use particular tracers (e.g.~RR Lyra stars) of the halo for which distances can be reliably estimated.  However, due to stellar evolution the tracers do not sample all possible ages and metallicities.  The advent of Gaia and the possibility to derive accurate distances for typical main sequence stars now affords the opportunity to test the degree to which tracer-based studies are biased.} \citep{belokurov2019}. In terms of the outer slope, some studies also find a steeper decline in the density profile at large radii, with a slope of $\approx -6$ outside $\approx 50$~kpc \citep{ivezic2000,deason2014}. However, other studies of outer regions ($\geq 50$~kpc) of the halo find that the density profile extends out to $\approx 100$~kpc without an obvious cut off \citep{fukushima2019} and with shalower slopes \citep{thomas2018,starkenburg2019}.  

M31 has a more metal-rich bulge/inner halo than the Milky Way, which is well fitted by an exponential profile. Outside this small region, the M31 halo is well fitted by a single power law. Various studies find similar values for the slope of the density profile, ranging between $\approx -2.75$ and $-3.3$ (see \citealt{gilbert2012} and references therein). We note also that observations have focused mainly on the metal-poor component of the M31 halo and/or on the minor axis, and therefore may underestimate the contribution of in situ stars (which are generally more metal-rich and near the disc). Nevertheless, the measured slope is generally consistent with a halo containing a high fraction of in situ stars, according to our simulations. Moreover, M31's profile does not display an obvious cut off out to at least a distance of $\approx 175$~kpc \citep{gilbert2012}. Such cut-offs appear commonly in accretion-only models \citep{bullock2005}. In contrast, haloes simulated hydrodynamically (and with high resolution) extend to $\approx 200$~kpc, without a sharp decline (see Fig.~\ref{fig:densprof}).  Understanding the origin of this apparent difference between self-consistent simulations and those based on accretion-only simulations would be helpful.

\subsubsection{Surface brightness profiles}

To facilitate the comparison with observations of external galaxies, we also analyse the $V$- and $r$-band surface brightness profiles of our simulated galaxies. Here we include all stars in the ARTEMIS haloes, as in general observational analyses do not attempt to separate different components but instead focus on radii where the disc and bulge are not expected to contribute significantly.  Since some observational data fields are preferentially positioned along the major and minor axes of galaxies (e.g.~the GHOSTS survey), we also analyse the simulated profiles along these axes.  To do so, we orient the simulated galaxies so that the discs are edge-on and select star particles in slabs of width of $10$~kpc along the minor and major axes.  

\begin{figure*}
  \includegraphics[width=\columnwidth]{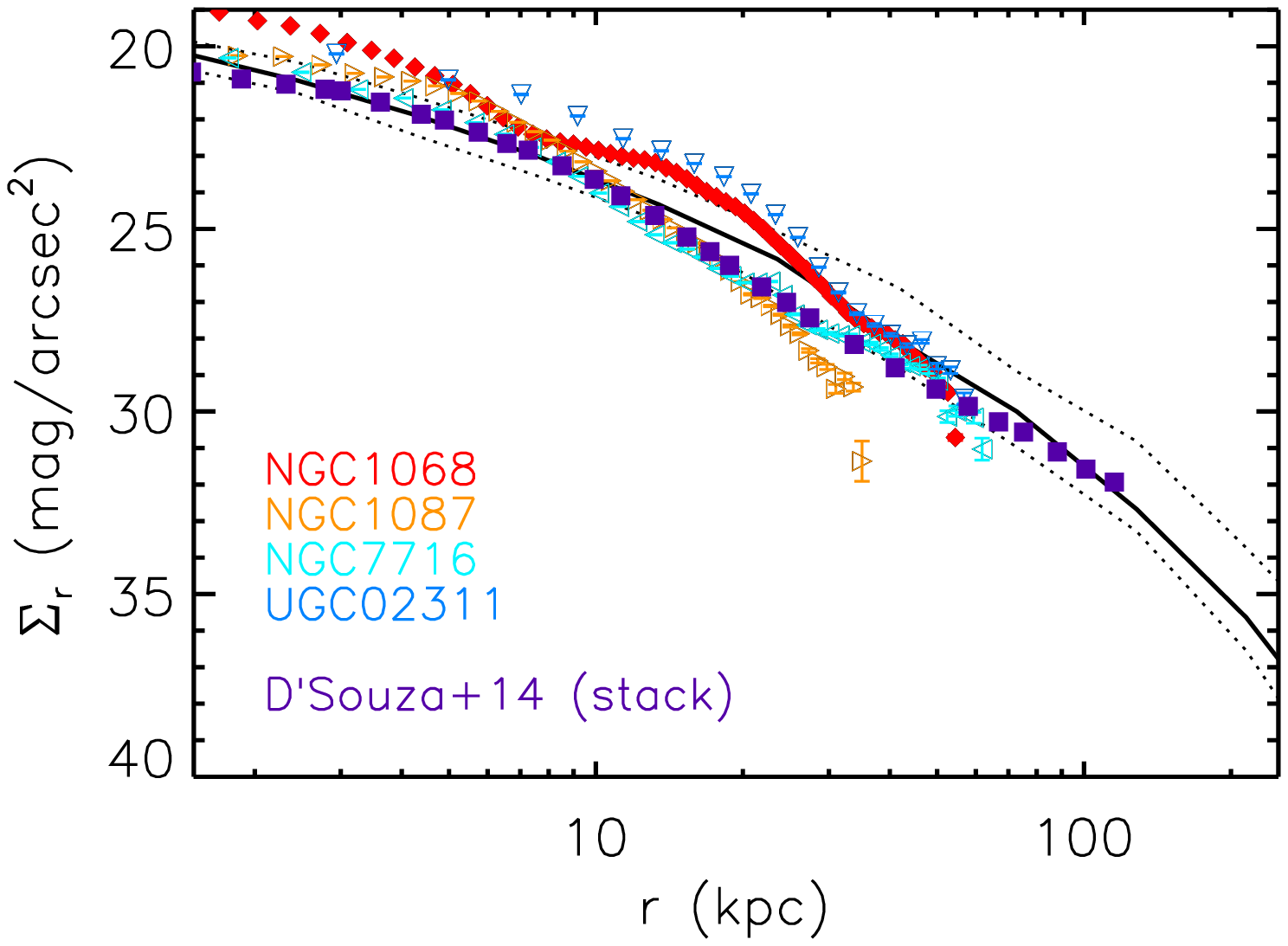}
  \includegraphics[width=\columnwidth]{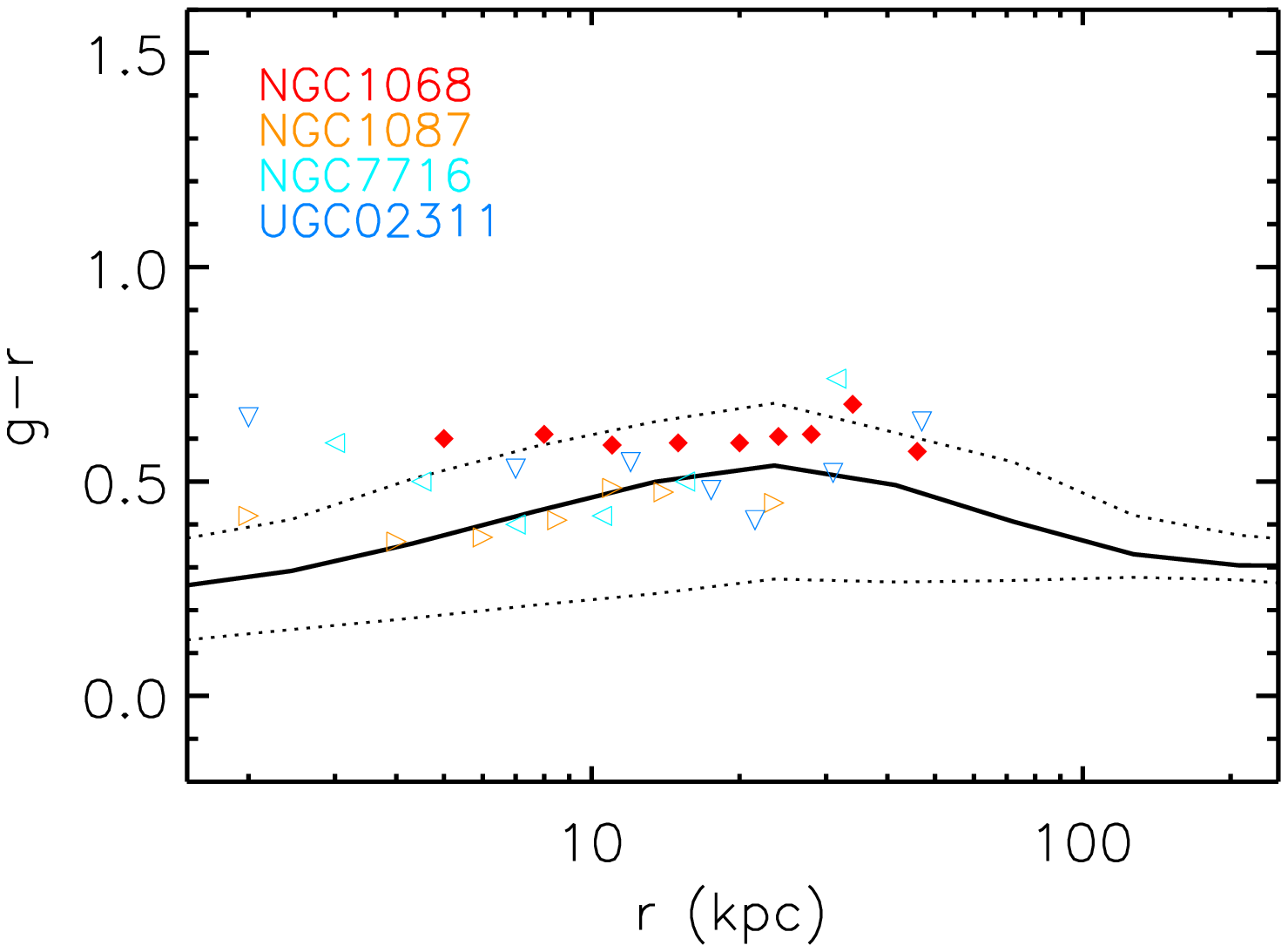}
  \caption{{\it Left:} The median r-band surface brightness profile, $\Sigma_{r} (r)$, of the simulated galaxies (solid black curve) and 1-sigma scatter ($16^{\rm th}$ and $84^{\rm th}$ percentiles, dotted black curves). Various coloured data points are $\Sigma_{r}$ measurements in $4$ disc galaxies from the sample of \citet{bakos2012} that fall within our simulated mass range. All galaxies are face-on.  Also plotted is the stacked SDSS $r$-band profile of \citet{d'souza2014} for Milky Way-mass galaxies. {\it Right:} The median simulated $g$-$r$ profile and its scatter (purple curves) versus $g$-$r$ data from \citet{bakos2012} (symbols are as in the left panel).}
    \label{fig:surfr_gr_data}
\end{figure*}

Fig.~\ref{fig:surfV_data} shows the $\Sigma_V(r)$ profiles along the minor axes (left panel) and along the major axes (right panel), respectively. The solid black curve shows the median profile of the ARTEMIS systems, while the dotted black curves show the corresponding scatter ($16^{\rm th}$ and $84^{\rm th}$ percentiles).  The various data points represent the observed fields from \citet{harmsen2017} along the minor and major axes of six edge-on or highly inclined disc galaxies in the GHOSTS survey (NGC253, NGC3031, NGC4565, NGC4945, NGC714 and NGC891), as well as measurements for M31 along both the minor axis \citep{gilbert2012,ibata2014} and major axis \citep{ibata2014}.  Note that the six galaxies from the GHOSTS survey have stellar masses between $(4$ - $8)\times10^{10}\, {\rm M}_{\odot}$, similar to the mass of the Milky Way, though are somewhat more massive than the median ARTEMIS system.
 
The simulated $\Sigma_V(r)$ profiles are generally in excellent agreement with the observations spanning a decade in radius.  We highlight here that this is {\it not} a result of calibration of feedback, which was only adjusted to approximately reproduce the integrated stellar mass within 30 kpc (most of which is within the central $\approx5$ kpc, corresponding to the half-mass radius).  

A clear exception is M31's minor axis, particularly in the SE direction.  In Fig.~\ref{fig:surfV_data} (left panel) we show observational measurements from the SE region (filled red circles are data from \citealt{gilbert2012} and filled red stars are those from \citealt{ibata2014}) and in the NW region (empty red stars are data from \citealt{ibata2014}).  Note that the SE region contains the Giant Stream, a very bright, metal-rich component, the progenitor of which has yet to be identified.  In contrast, the NW region is less contaminated by recent mergers and, consequently, the minor axis profile is in better agreement with the median simulated profile and with those of GHOSTS galaxies. Along the major axes (right panel), the simulated profiles agree very well with those of GHOSTS galaxies and of M31.  Furthermore, the simulations also appear to capture the break in the light profiles (particularly in the minor axis), as well as the slope in the outer regions, along both the minor and the major axes. 

Note that both simulations and observations are generally better fit by a BPL than by a SPL. The simulated $\Sigma(r)$ profiles can be fit essentially with the same BPLs as the $\rho_*(r)$ profiles, but with slopes given by $\alpha=\gamma+1$ (this neglects possible variations of age and metallicity with radius, which are explored below). The six GHOSTS galaxies may also be better fitted by BPLs (this can be inferred from a visual inspection of Fig.~\ref{fig:surfV_data}; see also \citealt{harmsen2017}). However, for simplicity, \citet{harmsen2017} fitted their data using an SPL. In M31, the surface brightness profile along the minor axis does not show an obvious downward break, so the data were also fitted by an SPL. Therefore, for consistency, we fit the outer parts of the simulated profiles with an SPL of the form $\Sigma_V \propto r^{\alpha}$,  where (projected)  $r >25$~kpc. The best SPL fits to the median simulated profiles are shown in Fig.~\ref{fig:surfV_data} with dotted curves. The slopes of these fits are $ \alpha_{\rm minor} = -2.21$ and $\alpha_{\rm major} = -2.33$ for the minor and major axes, respectively.  

These results indicate that the $\Sigma_V$ profiles along the major axes are generally somewhat steeper than those along the minor axes, which is also what is observed. For example, \citet{harmsen2017} have found best-fit values for the major axes slopes in the range of  $-5.33$ to $-2.73$ versus $-3.71$ to $-2$ for the minor axes.   The slope along the minor axis of M31 is $\simeq -2.2$ \citep{gilbert2012}, which is in agreement with the minor axis slope for the simulated galaxies ($\simeq -2.21$). 

In addition, we find that the major axes contain more light than the minor axes. This can be seen in Fig.~\ref{fig:surfV_data} for both simulated profiles and observations (at a fixed projected radius, major axes have lower $\Sigma_V$, i.e., are brighter, than the minor axes). These differences can be explained by the different stellar content along these two directions. The major axes contain both disc stars and halo stars passing near the disc plane. The halo stars could be accreted, but - as our simulations suggest - a large fraction also formed in situ.  Minor axes, however, are expected to contain mainly accreted stars, as stars ejected from the disc do not reach large heights. The different composition of accreted vs.~in situ stars along the two axes is reflected also in the different outer slopes. This is evident even in observations that may somewhat underestimate the contribution of in situ stars (for example in GHOSTS, where fields were chosen either along the minor axes or at relatively large distances along the major axes).

\begin{figure*}
    \includegraphics[width=\columnwidth]{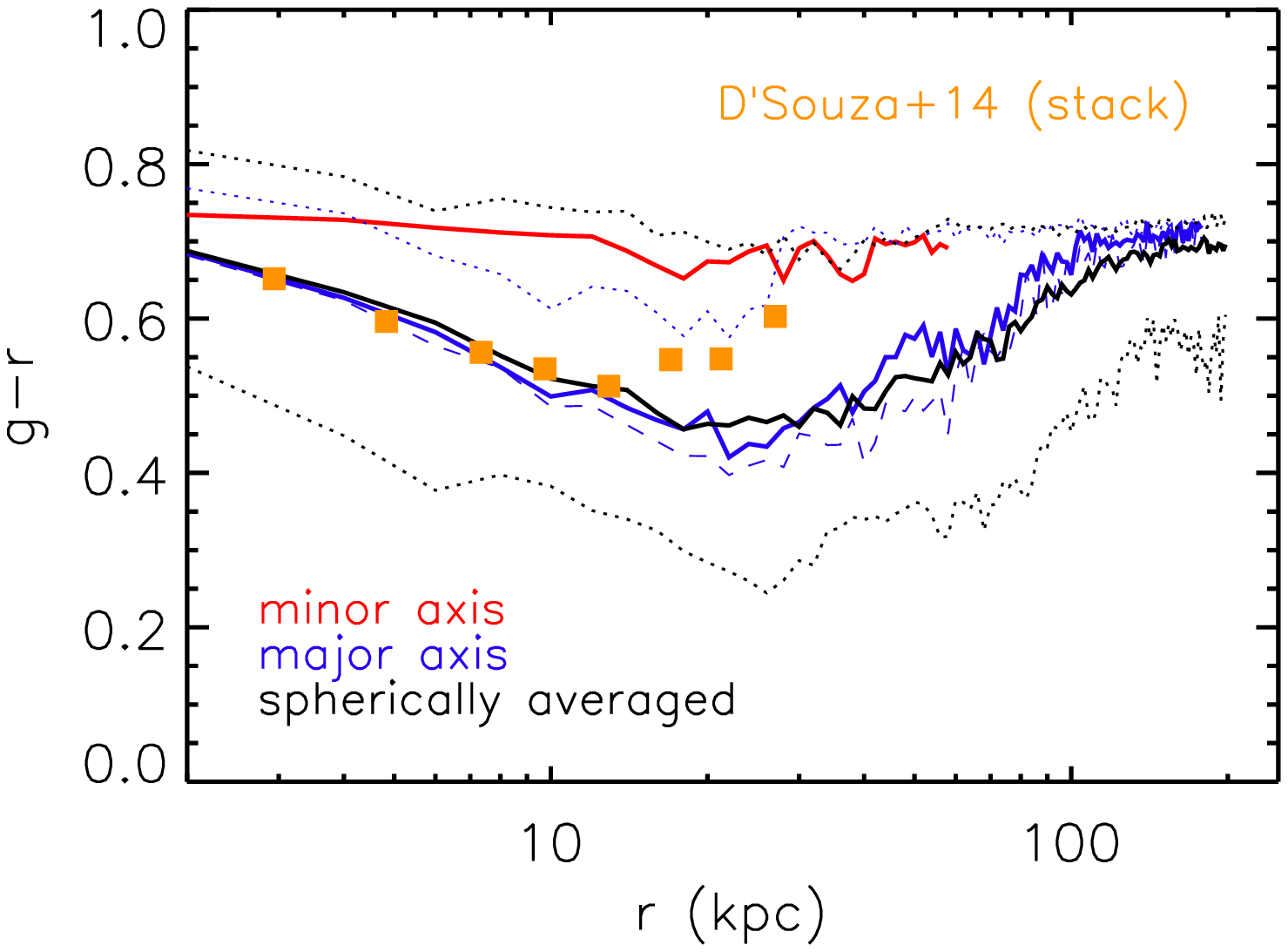} 
    \includegraphics[width=\columnwidth]{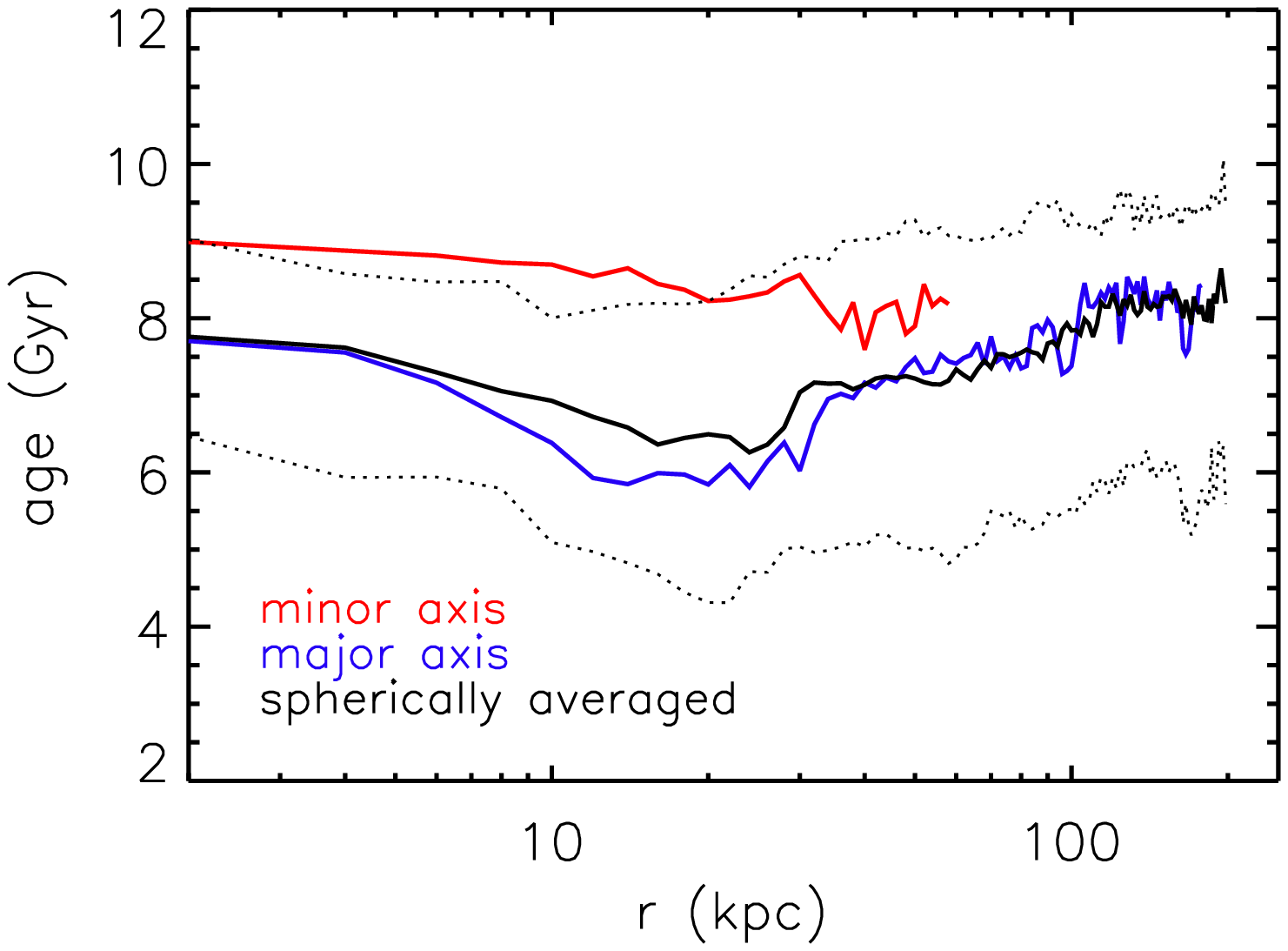} 
    \caption{{\it Left:} A comparison between simulated $g-r$ profiles and observations. The solid coloured curves represent the median colour along the major axis (blue), along the minor axes (red) and spherically-averaged (black).  The dotted blue and dashed blue curves represent the median major axis profiles for the accreted and in situ components, respectively.  The black dotted curves show the corresponding galaxy-to-galaxy scatter ($16^{\rm th}$ and $84^{\rm th}$ percentiles) for the spherically-averaged case. The filled orange squares show the azimutally-averaged $g-r$ colour of Milky Way-mass galaxies in SDSS DR9 from \citet{d'souza2014}. {\it Right panel:} The median age profile of the simulated galaxies. The coloured curves have the same meaning as in the left panel.}
    \label{fig:colour_age}
\end{figure*}

Fig.~\ref{fig:surfr_gr_data} (left panel) shows a comparison between the $r$-band surface brightness profiles, 
$\Sigma_r(r)$, from the simulations (the median profile) with the observations of \citet{bakos2012} which use integrated light to measure the haloes of several face-on disc galaxies. Here we use $4$ of the $7$ galaxies in their sample which fall within the luminosity/stellar mass limit of the simulated sample: NGC1068, NGC1087, NGC7716 and UGC02311.  (The excluded galaxies, NGC0450, NGC0941, UGC02081, are of considerably lower lumiosity/stellar mass than the Milky Way.) For this comparison, we orient the simulated galaxies so that the discs are face-on. This figure shows, again, very good agreement with the observations. The observed light profiles display breaks around $10$-$20$~kpc,  which plausibly demarcates the transition from the disc to the stellar halo component (e.g., \citealt{trujillo2020}; note that the curves are total surface brightness profiles, not just the bulge+halo component).  We also show the stacked SDSS DR9 $r$-band profile of \citet{d'souza2014} for Milky Way-mass galaxies (selected purely on stellar mass with no morphological criterion, as in the case of ARTEMIS), which is in excellent agreement with the simulations.

\subsubsection{Colour and age profiles}
\label{sec:colour_age}

In the right panel of Fig.~\ref{fig:surfr_gr_data} we compare the median colour ($g$-$r$) radial profile of the ARTEMIS systems with the observed profiles from the \citet{bakos2012} sample. As described by \citet{bakos2012}, the observed haloes display an up-turn in $g$-$r$ (haloes become redder around $10$~kpc). The simulations in Fig.~\ref{fig:surfr_gr_data} suggest that the reddest colours are expected roughly near the breaks in the $\Sigma_r(r)$ profiles. These are regions dominated by the disc and/or the in situ halo stars, which are mostly metal-rich. At large radii ($>50$~kpc), the simulated haloes contain mostly (metal-poor) accreted stars, hence the somewhat bluer colours. However, the relation between colour and metallicity is not linear, as the ages of stars also play a factor. This will be investigated below.

In Fig.~\ref{fig:colour_age}, we shift again to the edge-on view and study the $g$-$r$ profiles along the minor and major axes.  The left panel of Fig.~\ref{fig:colour_age} shows the median $g$-$r$ colour profile of the ARTEMIS systems, either along the major axis (blue curves) or the minor axis (red curves), as well as for the spherically-averaged case (black curves).  The dotted and dashed blue curves show the contributions of the accreted and in situ components (respectively) to the major axis colour profile.

The colour profiles along the major axis and that of the spherically-averaged case are very similar and are dominated by the in situ component out to large radii (compare the dashed and solid blue curves).  Both of these profiles display a similar up-turn in the $g$-$r$ colours, as observed in the sample of galaxies of \citet{bakos2012}.  With orange square symbols we also show the stacked extinction-corrected $g$-$r$ colours of Milky Way-analog galaxies in SDSS DR9 from \citet{d'souza2014}. 
The simulated (spherically-averaged) colour profiles are in generally good agreement with the SDSS data, including the up-turn around $r \approx 10$--$20$~kpc. Similar to the break in the r-band profiles in Fig.~\ref{fig:surfr_gr_data}, the up-turn in colours is plausibly the result of a transition from a disc-dominated region (e.g., \citealt{trujillo2020}) to the in situ-dominated stellar halo.  The up-turn happens at slightly larger radii in the simulations, which may be due to the simulations having somewhat flatter metallicity gradients compared to the observations (see Fig.~\ref{fig:fehprof} below).

The $g$-$r$ colour profile along the minor axis is relatively flat and consistently redder than the one along the major axis.  This is because the minor axis is dominated by accreted stars, originating from dwarf galaxies containing stellar populations with similar metallicities and older stellar populations.  (Note also that the accreted component of the major axis, represented by the dotted blue curve, is very similar to the total minor axis profile.) The relatively homogeneous mix of metallicities in the accreted population is one factor which leads to the lack of a significant colour gradient along the minor axes. This result is in general agreement with the measurements in some galaxies from the GHOSTS sample, which also show no significant colour/metallicity gradients (see Fig.~\ref{fig:fehprof} below) along the minor axes \citep{monachesi2016a}.  

In the outer regions ($r>50$~kpc), the colours along the minor and major axes begin to converge towards the red end.  Outer haloes are mostly formed through accretion and the stellar populations here are expected to be old (even though they were recently accreted).

\begin{figure*}
  \includegraphics[width=\columnwidth]{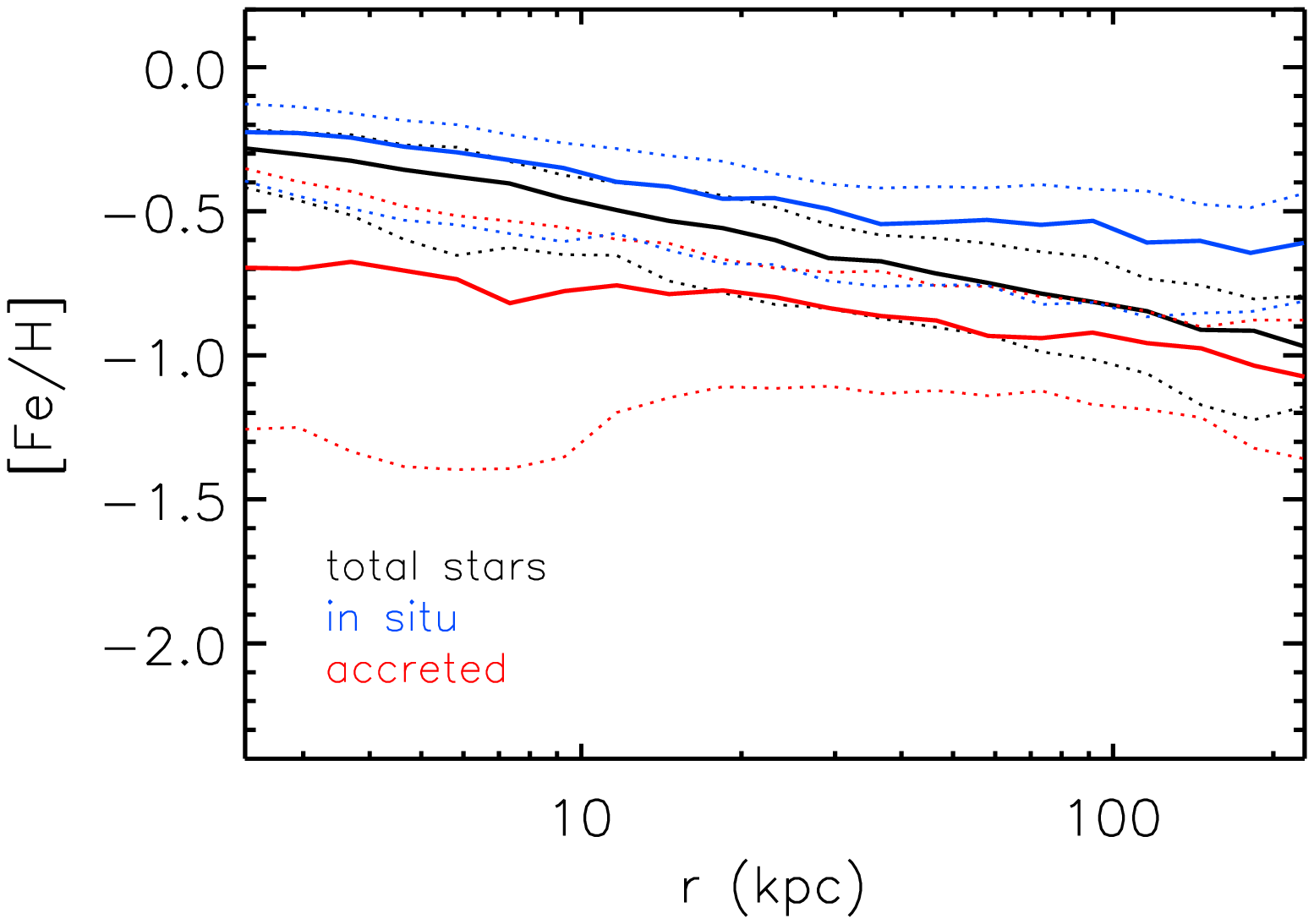}
  \includegraphics[width=\columnwidth]{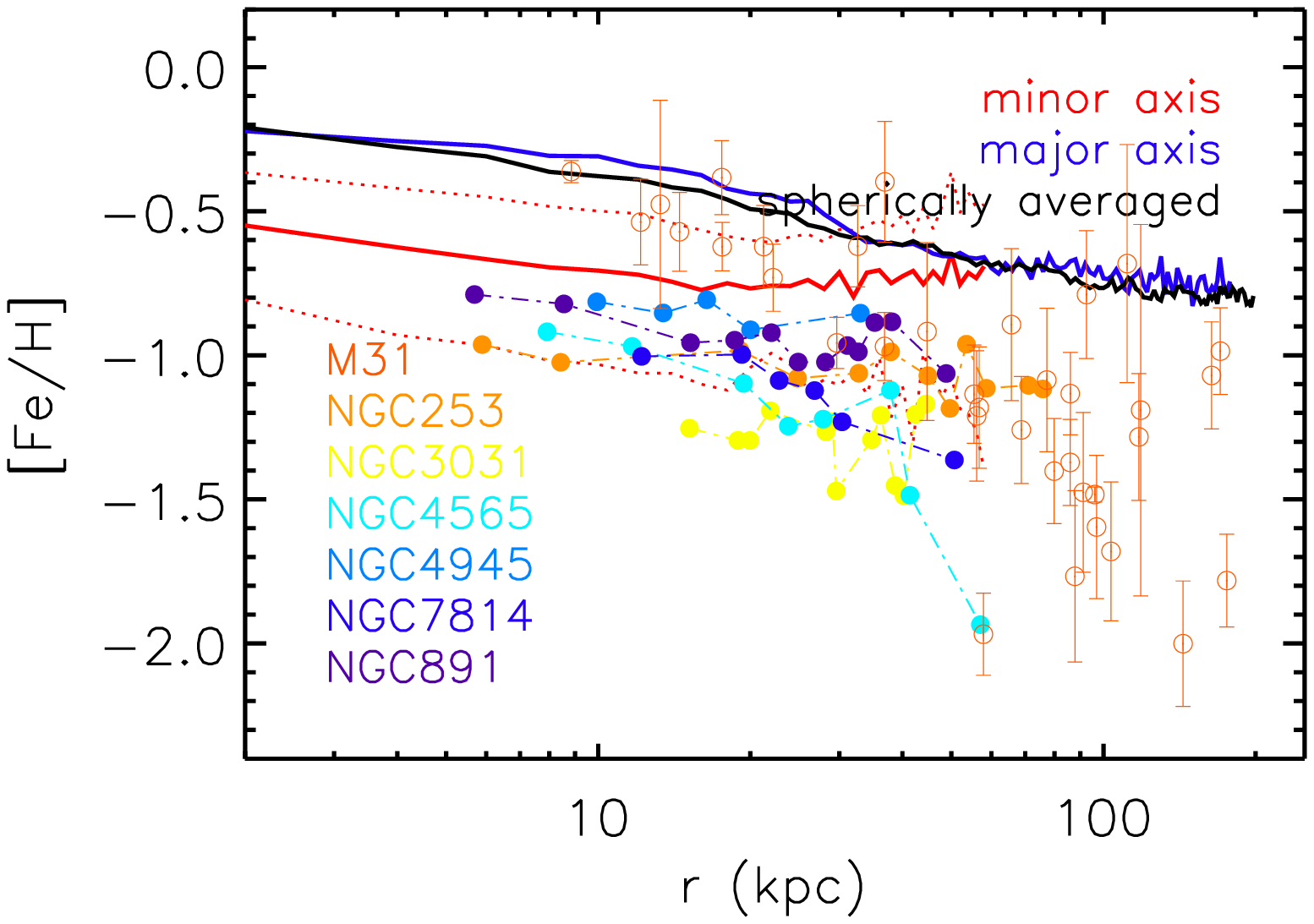}
  \caption{Median [Fe/H] profiles on the simulations. {\it Left:} Median [Fe/H] profiles of stars in the simulations: the total stellar component is shown with black curves, the in situ component in blue, while the accreted component in red. The full curves represent the median of each respective component and the dotted curves show the scatter around the median. {\it Right:}  Comparison between the simulated [Fe/H] profiles and observations. The coloured curves represent the [Fe/H] along the major axis (blue), along the minor axes (red) and spherically-averaged (black). The empty orange circles show the observational data in the M31 halo from \citet{gilbert2014} and the filled circles show the data in the haloes of six GHOSTS galaxies from \citet{monachesi2016a}.}
    \label{fig:fehprof}
\end{figure*}

In the right panel of Fig.~\ref{fig:colour_age} we show the median age profiles in the simulations, along the major and minor axes and spherically-averaged. To derive the median profiles, we compute the mass-weighted age profiles of each ARTEMIS system and then take the median of the profiles.  The age profiles display a strong similarity to the colour profiles. This shows that the red populations in the inner regions (i.e., along the minor axis) are also older (ages $\approx 8$-$9$~Gyr). The bluer populations along the major axes have younger/intermediate ages ($\approx 6$-$7$~Gyr), tracing the more recent episodes of star formation in the disc. The outer parts of the disc and the halo contain older ($\approx 8$~Gyr) populations brought in by accretion of low-mass satellite galaxies (see also \citealt{ruiz-lara2016}). Note, though, that the simulations predict a large scatter for  both the colours and ages of stellar populations.

As both the age and metallicity are important in determining the colour of a stellar population, it can be difficult to disentangle these quantities from the colour alone.  With the simulations, we have the advantage of being able to calculate the $g$-$r$ and [Fe/H] independently: while $g$-$r$ is derived from the light properties calculated with the methods described in Section \ref{sec:mainprops}, [Fe/H] traces the metal enrichment directly. Below we investigate the simulated radial [Fe/H] profiles.

\subsubsection{Metallicity profiles}
\label{sec:met}

Fig.~\ref{fig:fehprof} shows the median [Fe/H] profiles of the ARTEMIS systems, analysed in terms of the origin of the stars, i.e., in situ versus accreted stars (left panel), or measured along the minor and major axes (right panel).  We find that the in situ stars are consistently more metal-rich than the accreted stars, typically by $\approx 0.5$~dex, although this can be higher as the accreted stars have a larger [Fe/H] scatter. Both the in situ and accreted components display mild metallicity gradients. For accreted stars, the [Fe/H] gradient is less than $\approx 0.5$~dex over the scale of the halo, which is consistent with results  of accretion-only simulations \citep{font2006,cooper2010}. The in situ metallicity gradient is also weak,  $\approx 0.5$~dex over the scale of the halo. Overall, the total [Fe/H] profile is slighty steeper, because the two populations have distinct metallicities (in situ stars are consistently more metal-rich) and occupy different regions in the galaxy (in situ stars are more centrally concentrated). 
The right panel of Fig.~\ref{fig:fehprof} shows that the stars along the major axis are, typically, more metal-rich than those on the minor axis. Comparing with the left panel of the same figure, we can infer that stars along the major axis are preferentially born in situ. This includes, of course, the disc stars; however, given that the simulated stellar haloes are flattened (see Section \ref{sec:densprof}), the major axes probe a significant fraction of halo stars. This suggests that observations that target preferentially the minor axes may probe mainly the accreted halo. If stellar haloes contain a high fraction of in situ stars, as predicted by the simulations, then this type of observation may miss the bulk of the halo.

The gradient is more evident along the major axis or when metallicities are spherically-averaged (blue and black curves, respectively). In ARTEMIS, the median [Fe/H] gradient is $\approx 0.65$~dex over a distance of $\approx 150$~kpc. This does not appear sufficient to explain the strong [Fe/H] gradient detected in the M31 halo (empty orange circles; \citealt{gilbert2014}), although it is not clear whether M31 displays a gradient typical for its mass.  In the simulations, we expect the strength of the gradient will depend to some extent on the implementation of stellar feedback.  As we discussed in Section \ref{sec:scaling_rel}, the simulated dwarf galaxies are more metal-rich than observed\footnote{As a caveat, we have only compared the simulations to the observed mass--metallicity relation at $z=0$.  Since this relation evolves with time, and the accreted stellar halo as assembled over long timescale, the resulting metallicity gradient will also depend on the evolution of the mass--metallicity relation.}. This may result in too metal-rich outer haloes (which are built from accreted material) and hence to an overall shallower [Fe/H] gradient. In Section \ref{sec:compare_sims} we compare simulations with different feedback schemes and show how this may affect the [Fe/H] gradients.  

Along the minor axis, the simulated metallicity gradient is more shallow (solid red curve), although there is a considerable scatter among the simulated galaxies (see dotted red curves). These results are similar to those obtained with the Auriga simulations \citep{monachesi2016b}, who also obtain found significant differences between the [Fe/H] gradients along the two axes. Our results are also in agreement with observations of galaxies in the GHOSTS survey (filled coloured circles), which in general do not have significant colour or metallicity gradients along the minor axes \citep{monachesi2016a}.  Note, however, that typically the ARTEMIS stellar haloes along the minor axis are more metal-rich (i.e. an amplitude offset of $\approx0.4$~dex) than the GHOSTS galaxies.  Given that the minor axis preferentially probes the accreted component, this discrepancy may be expected given the offset in the mass--metallicity relation in Fig.~\ref{fig:props}.

\begin{figure*}
  \includegraphics[width=0.8\textwidth]{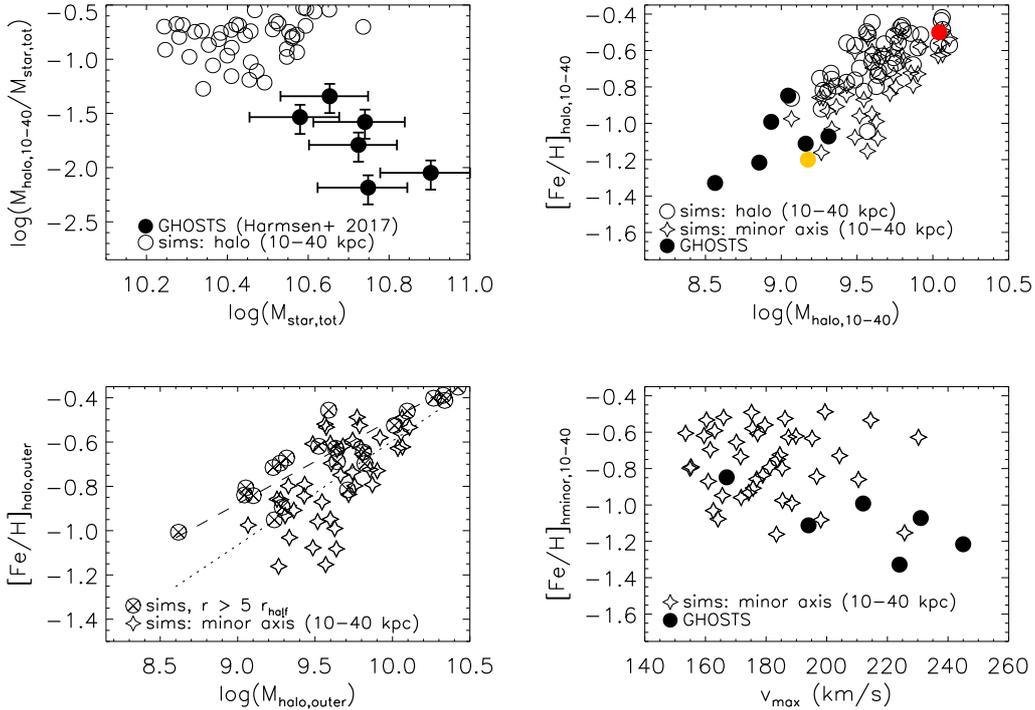}
    \caption{Relations between the properties of the outer stellar haloes and those of galaxies. {\it Top left:} Stellar mass fractions in the outer haloes, ${\rm log}({\rm M}_{\rm halo, \, 10-40}/{\rm M}_{\rm star, \,tot}),$ versus the total stellar masses of galaxies, ${\rm log}({\rm M}_{\rm star, \,tot})$. Simulations are shown with empty circles and GHOSTS galaxies with filled circles. {\it Top right:} The  metallicity -- stellar mass relation for the outer haloes, [Fe/H]$_{\rm halo, \,10-40}$ -- ${\rm M}_{\rm halo, \,10-40}$. In simulations, [Fe/H] are measured within 10 - 40~kpc, both along the minor axis (empty stars) and within a shell (empty circles). Observations include measurements along the minor axes in GHOSTS galaxies (filled circles) and measurements in the haloes of the Milky Way (yellow circle) and M31 (red circle). {\it Bottom left:} The metallicity -- stellar mass relation for simulated outer haloes: [Fe/H]$_{\rm minor, \,10-40}$ -- ${\rm M}_{\rm halo, \,10-40}$ (empty stars) and [Fe/H]$_{\rm halo, outer}$ -- ${\rm M}_{\rm halo, outer}$ (crossed circles), where the outer halo is considered outside $5\, r_{\rm half}$. The corresponding best fits to these relations are shown with dotted and dashed lines, respectively. {\it Bottom right:} The relation between the halo metallicity along the minor axis, [Fe/H]$_{\rm halo, minor, \,10-40},$ and the maximum circular velocity of galaxies, ${\rm v}_{\rm max}$, for simulations and the GHOSTS observations.}
    \label{fig:halo_gal}
\end{figure*}

\subsection{Stellar halo scaling relations}
\label{sec:halo_scalings}

It is well established that the total stellar mass of galaxies is strongly correlated with the total dark matter halo mass.  Furthermore, galaxy stellar mass exhibits strong correlations with other properties, such as star formation rate, size, and metallicity (see Fig.~\ref{fig:props}).  As stellar haloes are built from the accretion of smaller galaxies and, according to hydrodynamical simulations, an in situ component associated with dynamical heating of the early galaxy, we might expect stellar haloes to obey scaling relations in their own right.  Indeed, recent observational studies have indicated that the properties of stellar haloes do correlate with each other, as well as with properties of the main galaxy, such as total stellar mass \citep{gilbert2009,deason2016,bell2017,harmsen2017}.  Here we investigate a number of scaling relations in ARTEMIS and make comparisons to available observations, particularly the GHOSTS sample.

To enable a meaningful comparison with the observations from the GHOSTS survey, we define the halo here as a region between a fixed physical scale of $10-40$~kpc. In Fig.~\ref{fig:halo_gal} we plot several stellar halo--galaxy relations suggested by \citealt{harmsen2017} (see their Fig.~12) using our simulations and compare with the observations.

The top left panel in Fig.~\ref{fig:halo_gal} shows the stellar mass fractions of the haloes, $f_{\rm halo}\equiv{\rm M}_{\rm halo,10-40}/{\rm M}_{\rm star, tot}$, versus the total stellar masses of galaxies, M$_{\rm star, tot}$.  Over the narrow range of masses sampled in ARTEMIS, we see no strong evidence for a correlation between the stellar halo mass fraction and the galaxy stellar mass (thus, the stellar halo mass, rather than the fraction, scales approximately linearly with total stellar mass).  The GHOSTS disc galaxies, which are of slightly higher median total stellar mass than ARTEMIS, appear to show an anti-correlation between $f_{\rm halo}$ and ${\rm M}_{\rm star, tot}$, although the limited sample size and considerable scatter make it difficult to assess the robustness of this apparent trend.  Regardless of the trend, the observationally-inferred stellar halo mass fractions (i.e., the amplitude) are clearly lower than predicted by ARTEMIS.  This is surprising given the excellent match of ARTEMIS to the GHOSTS surface brightness profiles, shown previously in Fig.~\ref{fig:surfV_data}.  One possible reason for this discrepancy is how the value of ${\rm M}_{\rm halo,10-40}$ is estimated observationally.  GHOSTS consists of a limited number of pointings along the (especially) minor and major axes of galaxies.  To estimate the total stellar mass within the quoted aperture, \citet{harmsen2017} used the accretion-only N-body simulations of \citet{bullock2005} to calibrate the volume correction to be applied when inferring the stellar halo mass within 10-40 kpc from a sparse number of fields.  We speculate that the applied correction may be biased low, as the simulations adopted lacked an in situ component, which we find is dominant at these radii.  In addition, the stellar disc may contribute somewhat to the total stellar mass within the 10-40 kpc spherical shell in the simulations which would also not be accounted for in the volume correction applied to the GHOSTS galaxies. Alternatively, or perhaps in addition to the above, corrections are also required to convert the GHOSTS RGB counts into a stellar mass, which may also involve relevant systematic errors.

\begin{figure*}
  \includegraphics[width=\columnwidth]{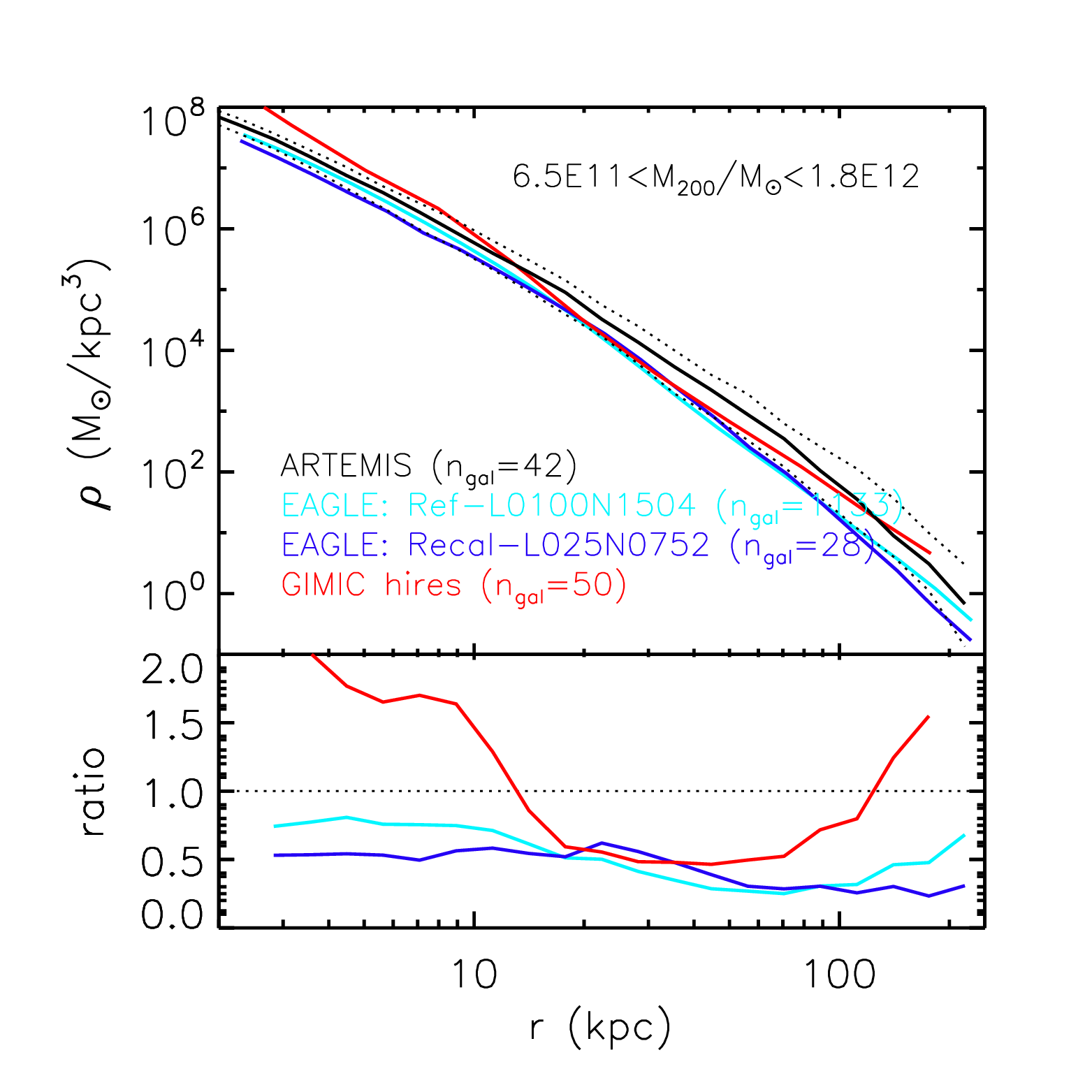}
  \includegraphics[width=\columnwidth]{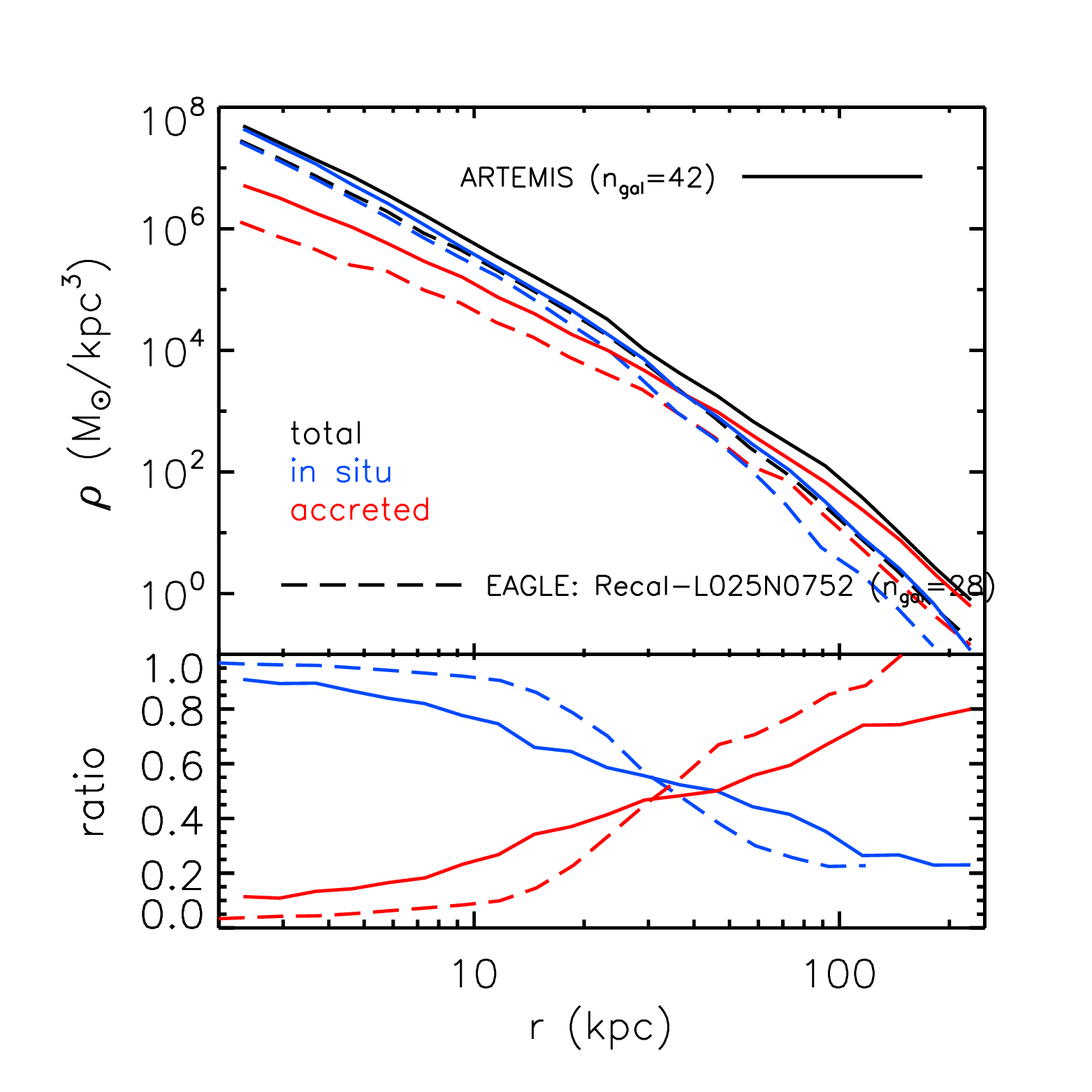}
  \caption{Comparison of the stellar mass density profiles predicted by different cosmological hydrodynamical simulations.  {\it Left:} Comparison of median total stellar mass density of Milky Way-analog haloes from ARTEMIS, the EAGLE Ref-L100N1504 model, the EAGLE Recal-025N0725 model, and the GIMIC `high-res' simulations. The number of galaxies in each sample, $n_{\rm gal}$ is shown in parentheses. Dotted black curves represent the scatter ($16^{\rm th}$ and $84^{\rm th}$ percentiles) about the median for ARTEMIS.  The subpanel shows the ratio of the other simulations with respect to the ARTEMIS result. {\it Right:} Comparison of the in situ and accreted contributions to the total stellar mass density between ARTEMIS and the EAGLE Recal-025N0725 model.  The subpanel shows the ratio of the in situ and accreted components with respect to the total from the respective simulations. }
    \label{fig:profs_eagle_rho}
\end{figure*}

The top right panel in Fig.~\ref{fig:halo_gal} shows the stellar halo metallicity, [Fe/H]$_{\rm halo, 10-40}$, versus stellar halo mass, M$_{\rm halo,10-40},$ for simulations and observations -- both indicating a clear correlation between these parameters. For the simulations, we measure [Fe/H]$_{\rm halo, 10-40}$ in two ways, within 10 - 40~kpc along the minor axes (empty stars) and in a spherical shell (empty circles), respectively.  Both are computed on a halo by halo basis.
Although these values are apparently similar in a visual inspection, quantitatively they are different, as discussed below. The observations include the haloes of six GHOSTS galaxies, for which [Fe/H]$_{10-40}$ have been measured along the minor axes (data from \citealt{harmsen2017} are shown with filled black circles); we also include the haloes of M31 (red circle) and of the Milky Way (yellow circle), for the latter using the revised values from \citet{deason2011} and \citet{conroy2019}. The simulated values lie comfortably within the bounds of the values obtained for the Milky Way and M31. They also agree with [Fe/H] measurements in GHOSTS galaxies.  However, given the discussion above regarding the possible bias in the observationally-inferred stellar mass estimates, the agreement in the amplitude of the [Fe/H]$_{\rm halo, 10-40}$--M$_{\rm halo,10-40}$ relation may be somewhat fortuitous.  If the observed stellar masses are indeed biased low, this would imply the simulations have higher halo [Fe/H] for a given (true) stellar halo mass, which would be consistent with the overall [Fe/H]--host galaxy stellar mass results shown in Fig.~\ref{fig:props} and the comparison of metallicity profiles in Fig.~\ref{fig:fehprof}.

With regards to the slope, we derive a slope for the simulated [Fe/H]$_{\rm halo, 10-40}$--M$_{\rm halo,10-40}$ relation of $\approx 0.46$, when the metallicities are measured along the minor axis (this best fit is shown in bottom left panel of Fig.~\ref{fig:halo_gal} with the dotted line). This slope is remarkably close to the value of $0.3$, predicted for haloes which are formed entirely through accretion, and assuming that the progenitor dwarf galaxies follow the stellar mass -- metallicity relation at present time \citep{dekel1986,kirby2013}. We expect, however, the bulk of the accreted halo to be assembled at an earlier time (although this may affect mainly the normalisation of the dwarf galaxy stellar mass -- metallicity relation, and not so much the slope). For GHOSTS galaxies, \citet{harmsen2017} obtain a slope of $0.7\pm0.15$. 

In the bottom left panel of Fig.~\ref{fig:halo_gal}, we also show the stellar halo metallicity--stellar halo mass and the associated best fit to this relation for the simulated galaxies using an alternative definition for the stellar haloes (e.g.~\citealt{merritt2016}), specifically as the region beyond $5 \, r_{\rm half}$ (arguably a more physically motivated definition than using a fixed physical scale along the minor axis, or even in a spherical shell). In this case, the best fit to the [Fe/H]$_{\rm halo, outer}$ -- M$_{\rm halo, outer}$ relation (shown with dashed line) returns a slope of $\approx 0.38$.  The shallower slope may suggest an additional contribution of in situ halo stars beyond $5 \, r_{\rm half} \simeq 25$~kpc, in agreement with the results obtained before. 

Finally, in the bottom right panel of Fig.~\ref{fig:halo_gal} we investigate the relation between the median [Fe/H] of the haloes, measured within $10$--$40$~kpc along the minor axes, and the maximum circular velocities of the galaxies, ${\rm v}_{\rm max}$. We compare these with the corresponding values of GHOSTS galaxies, as calculated by \citet{harmsen2017}. No clear relations can be found in either the observations or simulations between these two parameters. This suggests that the total mass of the galaxy (for which ${\rm v}_{\rm max}$ acts as a proxy) is not strongly correlated with the metallicity of the stellar halo (at least over this narrow range of total masses), unlike the halo stellar mass.  We also note that, at given ${\rm v}_{\rm max}$, the simulations have somewhat higher metallicities than observed, consistent with the discussion above.

\begin{figure*}
  \includegraphics[width=\columnwidth]{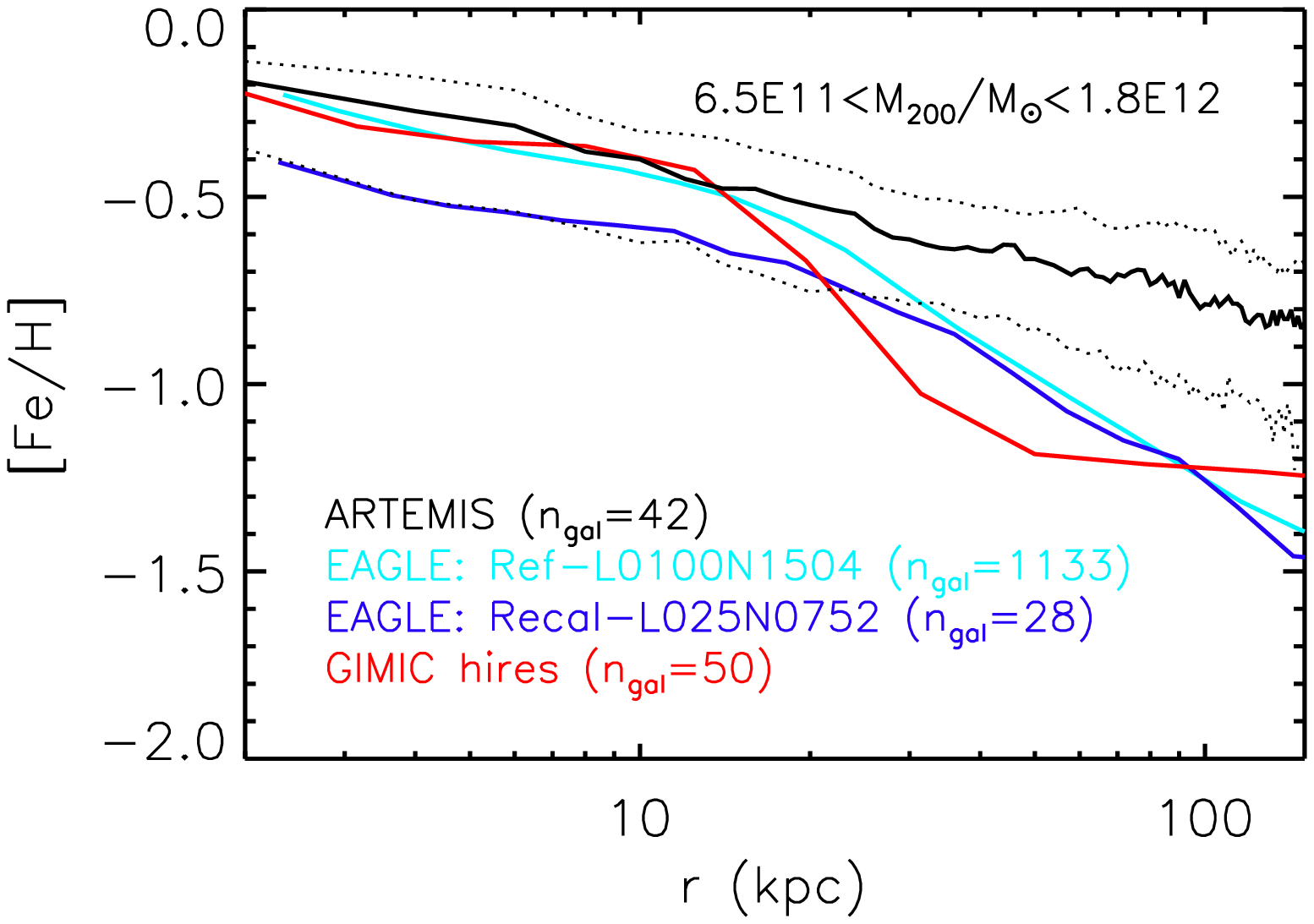}
  \includegraphics[width=\columnwidth]{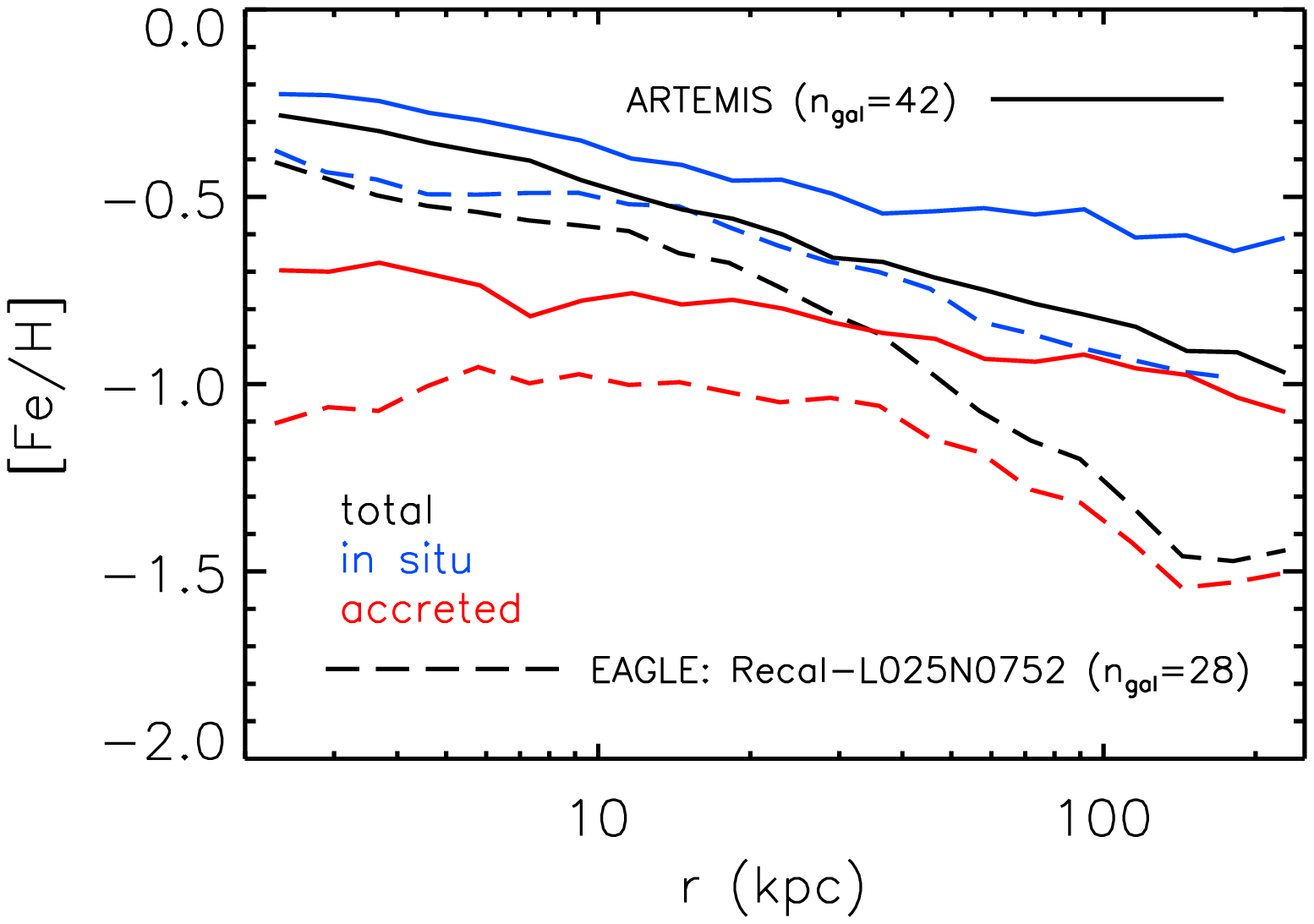}
  \caption{Comparison of the stellar metallicity profiles predicted by different cosmological hydrodynamical simulations.  {\it Left:} Comparison of median total median stellar metallicity profiles of Milky Way-analog haloes from ARTEMIS, the EAGLE Ref-L100N1504 model, the EAGLE Recal-025N0725 model, and the GIMIC `high-res' simulations. The number of galaxies in each sample, $n_{\rm gal}$ is shown in parentheses. Dotted black curves represent the scatter ($16^{\rm th}$ and $84^{\rm th}$ percentiles) about the median for ARTEMIS.  {\it Right:} Comparison of the in situ and accreted contributions to the stellar metallicity between ARTEMIS and the EAGLE Recal-025N0725 model.}
  \label{fig:profs_eagle_met}
\end{figure*}

\section{Comparisons with previous hydrodynamical simulations}
\label{sec:compare_sims}

How does ARTEMIS compare with other simulations in terms of stellar halo predictions?
Here we compare the predictions of ARTEMIS with those of EAGLE and GIMIC, two existing cosmological hydrodynamical simulations that match relatively well the properties of Milky Way-analog haloes (\citealt{schaye2015,crain2009}, respectively) and that also produce stellar haloes with a dual nature.  We choose to compare these particular simulations because their codes share many similarities, differing mainly in the scheme of stellar feedback. This allows us to focus on a more restrictive topic, specifically on how differences in the stellar feedback may affect the structure of stellar haloes, in particular, the contribution of in situ stars and the overall shape of the galaxy stellar density and metallicity profiles. A comprehensive comparison study involving a wider range of simulations would of course be very useful, but it is beyond the scope of this paper.

From the suite of EAGLE simulations, we use two models: the $100$~cMpc box Reference model, Ref-L0100N1504 (with gas particle mass of $1.81 \times 10^{6} \, {\rm M}_{\odot}$, and dark matter particle mass of $9.7 \times 10^6 \, {\rm M}_{\odot}$) and the $25$~cMpc box recalibrated model, Recal-L0025N0725, analysed earlier. From the GIMIC suite, we use only the highest resolution simulation (gas mass of $1.46 \times 10^6\, {\rm M}_{\odot}$ and dark matter mass $6.63 \times 10^6\, {\rm M}_{\odot}$). From each of these simulations we select galaxies in the same mass range as ARTEMIS, specifically with virial masses $6.5 \times 10^{11} {\rm M}_{\odot} < {\rm M}_{200} < 1.8 \times 10^{12} {\rm M}_{\odot}$. Within this mass range, there are $1133$ galaxies in EAGLE Ref-L0100N1504 and $28$ galaxies in EAGLE Recal-L0025N725. We also compare with the sample of $50$ Milky Way-halo mass galaxies from the GIMIC high-resolution simulation, respectively (see also \citealt{font2011} and \citealt{mccarthy2012a} for a more detailed analysis of the GIMIC sample). ARTEMIS has a comparable number of Milky Way-analog systems ($42$) as the highest resolution simulations in EAGLE and GIMIC, with the additional benefit that it has significantly higher numerical resolution, allowing us to resolve in greater detail the structure of stellar haloes. All these simulations assume a $\Lambda$CDM model, with small differences in the cosmological parameters that should not affect the properties of galaxies studied here.

As already mentioned, the subgrid physics in GIMIC and EAGLE is similar in many respects, but there are also a few notable differences.  Both simulations impose a pressure-density relation (an equation of state) for star-forming gas and the star formation prescriptions are similar, except that EAGLE uses a star formation threshold that is metallicity-dependent.  The main differences lie in the feedback implementations.  Specifically, GIMIC uses a kinetic implementation of stellar feedback (i.e., a set number of neighbouring gas particles are given a velocity kick), whereas in EAGLE the feedback is thermal, where the number of neighbouring gas particles heated is chosen stochastically given constraints on the fraction of available energy used for feedback and the adopted heating temperature.  Perhaps the most relevant difference, though, is that the stellar feedback was adjusted in EAGLE to match the galaxy stellar mass function and the size--mass relation (this was done by varying the dependence of stellar feedback energy on gas density and metallicity), whereas the stellar feedback in GIMIC was not adjusted to match a specific observational data set.  Another difference between GIMIC and EAGLE is that the former does not include feedback from AGN, although we do not expect that to be a particularly important omission for studies of stellar haloes around Milky Way-mass haloes.

Fig.~\ref{fig:profs_eagle_rho} shows a comparison between the median stellar density profiles $\rho(r)$ (left panel) and a comparison of the in situ and accreted component contributions to the total stellar density focusing here just on the ARTEMIS and EAGLE Recal models (right panel).  All profiles (shown with solid curves) are spherically-averaged.  We also plot the intrinsic scatter for the total ARTEMIS result (with dotted curves).  

The density profiles predicted by the three suites of simulations are in qualitative agreement with each other.  Differences are apparent in detail, though (see the ratios in the left subpanel).  In particular, ARTEMIS predicts higher stellar mass densities (by about a factor of 2) at all radii compared with the EAGLE simulations, which tallies with the fact that EAGLE lies somewhat below the abundance matching results for the stellar mass--total mass relation (see Fig.~\ref{fig:props}).  This in turn translates to a smaller contribution (in terms of stellar mass) from both the in situ component and the satellites that merge to form the accreted component.  GIMIC, on the other hand, predicts higher mass densities in the central regions ($r \la 10$ kpc) but slightly lower mass densities compared to ARTEMIS at larger radii (out to $\approx 100$ kpc).  Comparison with GIMIC is a bit more challenging, as GIMIC lies below the abundance matching results for stellar masses of $\log{M_{\rm star}} \la 10.5$ but lies above them at higher masses (see fig.~4 of \citealt{mccarthy2012b}).  In addition, the higher-mass galaxies in GIMIC are too compact with respect to observations.  It is plausible that the reason that GIMIC lies above ARTEMIS in the central regions is due to overcooling, while at large radii it lies below ARTEMIS because the accreted component in GIMIC was built from undermassive galaxies (that lie below the abundance matching results).

In the right panel of Fig.~\ref{fig:profs_eagle_rho} we plot the median stellar mass density profiles of ARTEMIS and of the EAGLE Recal-L0025N0752 run split according to in situ and accreted origins.  In agreement with ARTEMIS, the in situ component dominates the central $r \la 35$ kpc or so in the EAGLE Recal model.  Interestingly, even though the crossover points are similar, there are differences in the respective contributions interior and exterior to this point, which the accreted component playing a larger (smaller) role in the inner (outer) regions in ARTEMIS compared to the EAGLE Recal model.  In other words, there is a stronger segregation of in situ and accreted components in the EAGLE Recal model.

Turning now to the predicted metallicity profiles in the left panel of  Fig.~\ref{fig:profs_eagle_met}, all of the median [Fe/H] profiles display a gradient at some level. As discussed earlier, this behaviour is seen more clearly when the metallicities are spherically-averaged or when measured along the major axis.  However, there are significant differences between the predicted [Fe/H] radial profiles.  ARTEMIS shows the mildest [Fe/H] gradient, whereas GIMIC shows the strongest (see also \citealt{font2011}) out to $r\approx 40$ kpc, beyond which it flattens. One possible explanation for this difference is the numerical resolution.  Here the simulations with the highest resolution have the mildest gradients, and vice-versa.  However, a more likely explanation for the trends we see is that the simulations predict different relative contributions of the metal-rich in situ and metal-poor accreted components and that the stellar mass--metallicity relations of the host galaxies differ.  In particular, the steepness of the stellar mass--metallicity relation (which evolves with redshift) should effectively set the difference in the metallicities of the in situ and accreted components.  A steeper relation means that the satellites that build the accreted component will have a lower metallicity compared to that of the in situ component that originates from the central galaxy.  The mass fraction in the two components will also contribute to setting the steepness of the metallicity gradient.

To illustrate the above, in the right panel of Fig.~\ref{fig:profs_eagle_met} we plot the metallicity profiles of ARTEMIS and the EAGLE Recal run split according to in situ and accreted origins.  There is generally good agreement between the simulations in the sense that both the in situ and accreted components have mild individual gradients which combine to form a steeper metallicity gradient (owing the steeply changing in situ-to-accreted mass fraction with radius).  However, we see that the amplitude difference in metallicity of the two components is larger for the EAGLE Recal model than it is for ARTEMIS.  We ascribe this difference to the difference in the steepness of the stellar mass--metallicity relations. Furthermore, we have shown that there is a stronger separation of in situ and accreted components by mass fraction in the EAGLE Recal model compared to ARTEMIS (right panel of Fig.~\ref{fig:profs_eagle_rho}), which will act to steepen the gradient in that simulation.  The GIMIC simulations (not shown) are similar in behaviour to the EAGLE Recal model, in that they have a similarly large offset in the metallicities of the in situ and accreted components.  This, combined with the higher in situ-to-accreted mass fraction ratio in GIMIC (due to excessive star formation in the central regions, as discussed above) results in an even steeper metallicity gradient in GIMIC.

On the basis of the above, while ARTEMIS likely achieves a more accurate modelling of the stellar mass fractions in the two components than EAGLE or GIMIC, the predicted metallicity gradients from ARTEMIS are likely somewhat too shallow as a result of the stellar mass--metallicity relation being too shallow.  Overcoming this issue while not significantly affecting the stellar masses is non-trivial and likely requires feedback driven winds to be significantly more metal mass loaded.

\section{Conclusions}
\label{sec:concl}

We have introduced a new suite of high-resolution (baryon particle mass of $\approx 2.2 \times 10^4 {\rm M}_\odot/h$), cosmological hydrodynamical simulations of 42 Milky Way-analog haloes, called ARTEMIS.  These haloes were selected on the basis of halo mass, spanning the range $8\times10^{11} < {\rm M}_{200,{\rm crit}}/{\rm M}_\odot < 2\times10^{12}$ in a dark matter-only periodic box.  We have shown that these simulations match a variety of global and structural properties of the Milky Way and of other similar disc galaxies, although the predicted stellar mass-metallicity relation is somewhat too shallow. Focusing on the properties of simulated stellar haloes, we investigated the radial distributions of their stellar mass, surface brightness, stellar metallicity, stellar age and colour, and distinguished the signatures of the accreted and in situ stellar components in these profiles. We then performed detailed comparisons with the observations as well as with other cosmological hydrodynamical simulations.

Our conclusions can be summarised as follows:
\begin{itemize} 
\item The simulated stellar haloes have high fractions of in situ stars (Fig.~\ref{fig:densprof}): on average, in the solar neighborhood, $\approx 70\%$ of halo stars are predicted to have formed in situ, and the fractions remain high out to $30$--$40$~kpc.  This result is in good agreement with the Gaia-derived in situ halo fraction near the sun, which is estimated to be nearly $50\%$ \citep{belokurov2019}.

\item The simulated stellar density profiles are well fitted by broken power laws, with shallower slopes in the inner region ($\approx -3$) than in the outer region (slopes of $\approx -4$) - see Fig.~\ref{fig:slopes}. The break radii of these profiles are typically $\approx 20$--$40$~kpc, although they show large variations from halo to halo. The break radii overlap, and are indeed likely the result of, the transition between the in situ-dominated region and the accretion-dominated one. 

\item The simulated haloes show metallicity and colour gradients (though mild), particularly when the properties are spherically-averaged or when measured along the major axes of galaxies (Fig.~\ref{fig:fehprof}). In contrast, the metallicity and colour profiles along the minor axes are nearly flat, in agreement with the observations. This behaviour can be explained by the fact that the in situ component is highly flattened and aligned with the disc, whereas the accreted component is more isotropically distributed (on average).  Given the dominance of the in situ component, the total stellar halo is flattened within the central $\approx 30$ kpc, with $q \equiv c/a \approx 0.6$, in agreement with a variety of observations of stellar haloes, including the Milky Way \citep{sesar2011}, M31 \citep{ibata2005}, the Milky Way-analog galaxies in the GHOSTS survey \citep{harmsen2017}, and stacks of on edge-on galaxies in the SDSS \citep{zibetti2004}.

\item The simulated galaxies display higher stellar halo fractions than the expected values if haloes were to form via accretion-only. Simulations display a clear metallicity -- stellar mass relation for stellar haloes, however this relation has a shallower slope than the slope expected in an accretion-only scenario (Fig.~\ref{fig:halo_gal}).

\item Changes in the prescriptions for stellar feedback affect both the steepness of the stellar mass--metallicity relation of galaxies and the respective mass fractions in the in situ and accreted components.  These two characteristics are what determine the steepness of metallicity gradients in the stellar halo.  Previous simulations such as EAGLE and GIMIC predict somewhat steeper gradients than ARTEMIS, which we attribute to differences in these characteristics in the simulations (see Section \ref{sec:compare_sims}).  ARTEMIS achieves a better match to the stellar mass--halo mass relation and stellar mass--size relation compared to EAGLE and GIMIC.  To obtain steeper metallicity gradients within ARTEMIS, which appear to be present in the observations (at least of M31), would likely require invoking feedback driven winds that are preferentially metal mass-loaded.
\end{itemize}

Our study has addressed a number of important questions about the nature of stellar haloes and highlights the importance of modeling stellar haloes self-consistently (i.e., with hydrodynamics) in a full cosmological context.  The prominent in situ component predicted by the simulations is a rich repository of information about the formation history of galaxies. In the Milky Way, the in situ halo has long been eluding detection, likely due to limiting selection effects.  Until recently, observational studies have targeted stars that were either metal-poor or kinematically distinct from the disc. With the recent confirmation of the importance of the in situ halo in the Milky Way \citep{belokurov2019,conroy2019}, and with strong evidence in favour of an important in situ component of the M31 halo \citep{dorman2013}, there is a need to provide accurate predictions for the nature of stellar haloes using cosmological simulations. Our study has provided detailed information about the structure of in situ and accreted components and has made a number of predictions which can be tested in the future. Specifically, observations that target the major axes of galaxies or the disc/halo interface will be able to test the realism of our predictions.  Wide-field Galactic surveys such as Gaia or WEAVE can also produce samples of halo stars less affected by a-priori selection criteria.

\section*{Acknowledgments}
The authors thank the anonymous referee for constructive comments that improved the paper.  They also thank Vasily Belokurov and Ignacio Trujillo for helpful feedback.
This project has received funding from the European Research Council (ERC) under the European Union's Horizon 2020 research and innovation programme (grant agreement No 769130).  It was supported in part by the National Science Foundation under Grant No. NSF PHY-1748958.  This work used the DiRAC@Durham facility managed by the Institute for Computational Cosmology on behalf of the STFC DiRAC HPC Facility. The equipment was funded by BEIS capital funding via STFC capital grants ST/P002293/1, ST/R002371/1 and ST/S002502/1, Durham University and STFC operations grant ST/R000832/1. DiRAC is part of the National e-Infrastructure. A. Font acknowledges fruitful discussions on the nature of stellar haloes with colleagues at the KITP workshop ``Dynamical Models for Stars and Gas in Galaxies in the Gaia Era'' and at the Sexten CfA workshop "Light in the suburbs: structure and chemodynamics of galaxy halos".

\section*{Data availability}
The data underlying this article may be shared on reasonable request to the corresponding author.


\bibliographystyle{mnras}
\bibliography{references} 



\appendix
\section{Impact of metallicity definition on the stellar mass--metallicity relation}
\label{sec:appendixA}

\begin{figure}
  \includegraphics[width=\columnwidth]{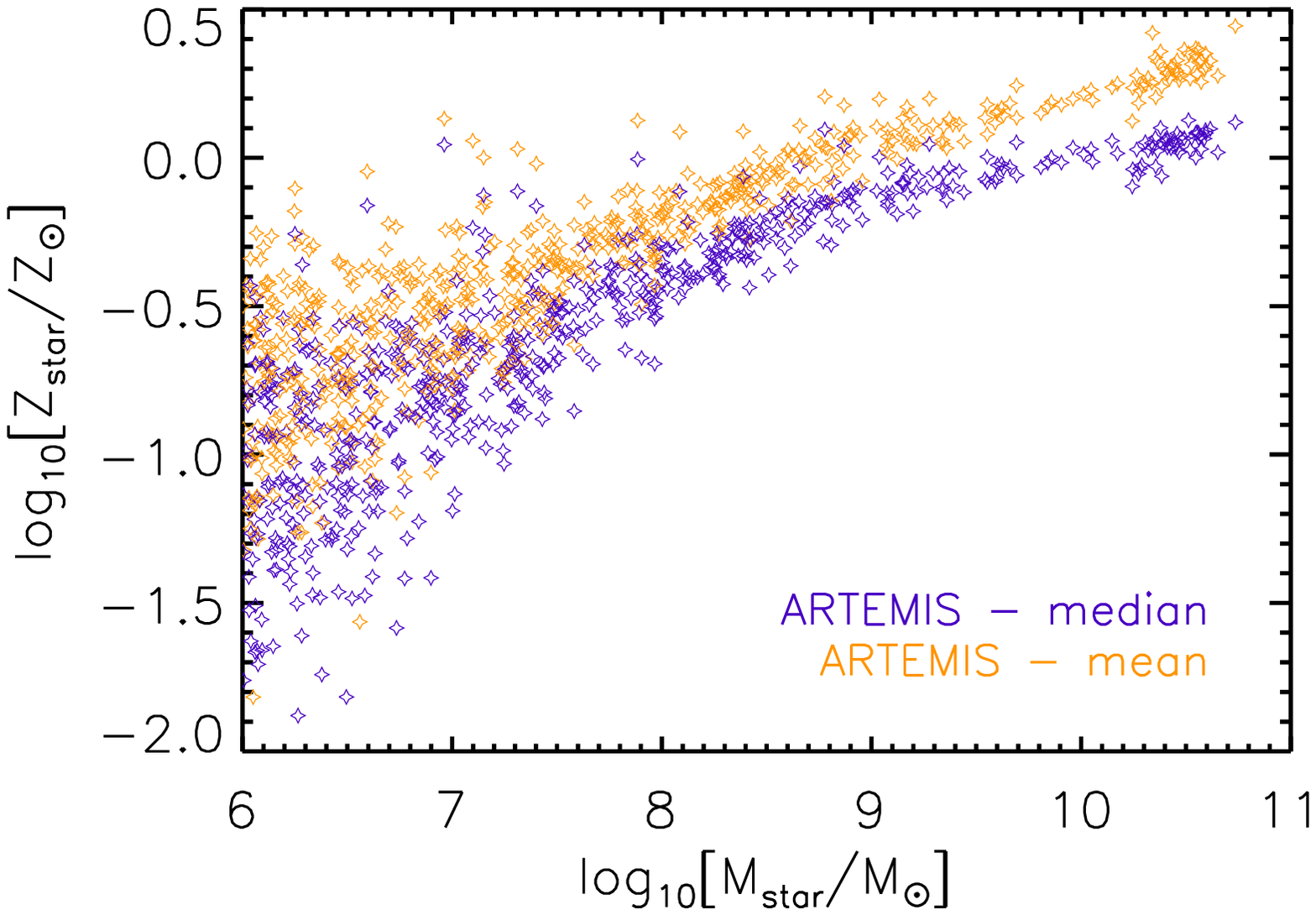}
    \caption{Comparison the stellar mass--metallicity relation of ARTEMIS using either the median star particle metallicity (empty purple stars) or the mass-weighted mean metallicity (empty orange stars), both for all stars within a $30$~kpc (physical) aperture.}
    \label{fig:zstar_med_mean}
\end{figure}

Here we explore the impact of metallicity definition on the stellar mass--metallicity relation.  In Fig.~\ref{fig:zstar_med_mean} we show the relations from ARTEMIS when the metallicity is defined as either the median star particle metallicity (empty purple stars) or the mass-weighted mean metallicity (empty orange stars), both for all stars within a $30$~kpc (physical) aperture.  At the high-mass end ($\sim 10^{10} \,{\rm M}_\odot)$, the mean metallicity is typically $0.2-0.3$~dex larger than the median with little scatter.  At lower masses, on the other hand, the difference ranges from $\approx 0.1-1$~dex, plausibly because the mean becomes much more sensitive to outliers due to the smaller number of star particles.

As discussed in the main text, which of these definitions is more appropriate for comparison with the observations is unclear.  The mean metallicity might be more appropriate for comparison with integrated light measurements, whereas the median may be more appropriate for observations of dwarf galaxies (and some observations of stellar haloes) where typically one measures the metallicities of a number of individual stars.

\section{Impact of varying morphological decomposition}
\label{sec:appendixB}

We explore here the impact of varying the threshold in $\kappa_{\rm rot}$ used to separate the disc from the bulge+halo component.  In our main analysis, we employ a threshold of $\kappa_{\rm rot} > 0.8$ to identify disc particles, following \citet{font2011}.  Here we show how the stellar bulge+halo density profiles change when the threshold is lowered to 0.5, implying that particles with at least half of their kinetic energies in ordered rotation are assigned to the disc.

\begin{figure*}
     \includegraphics[width=\columnwidth]{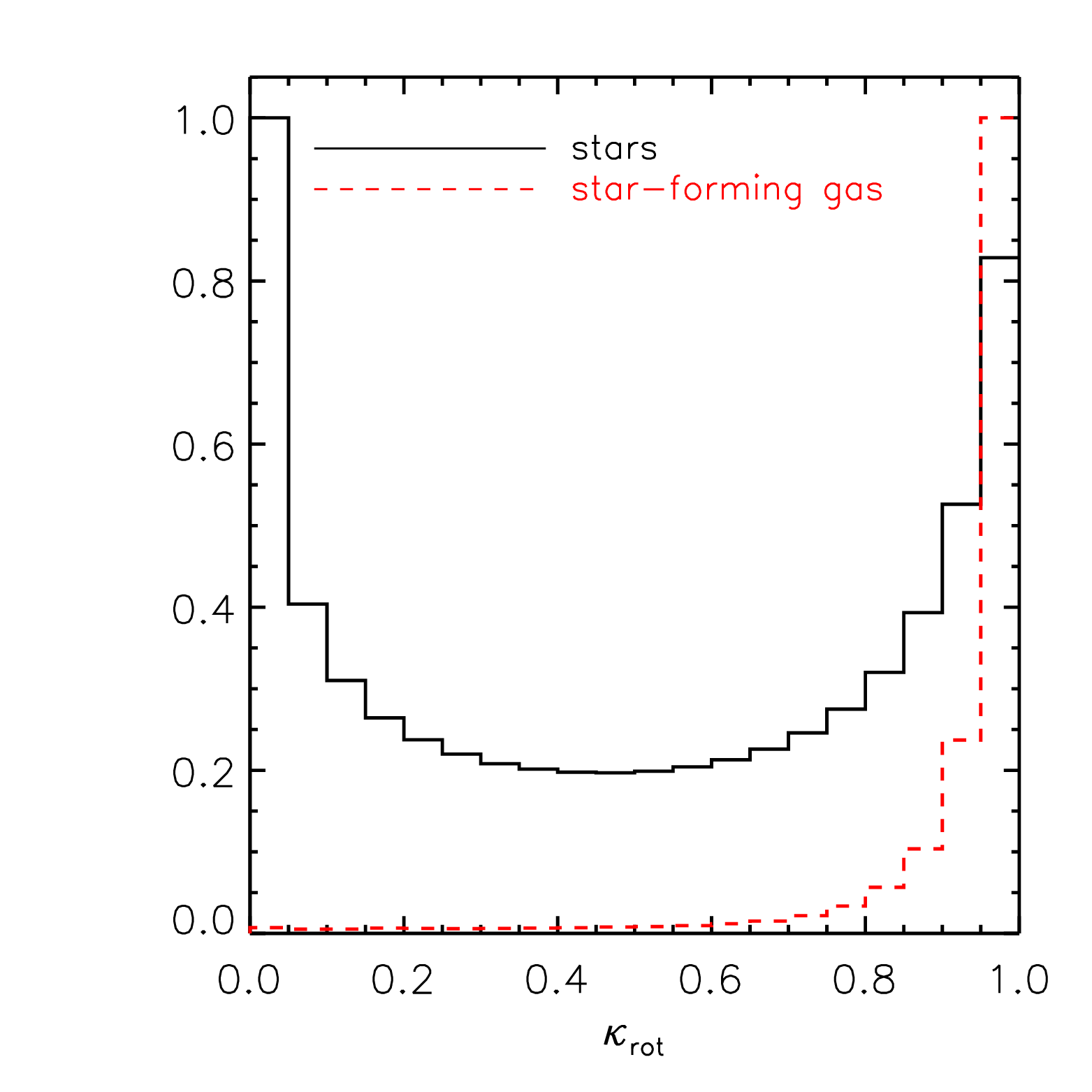}
    \includegraphics[width=\columnwidth]{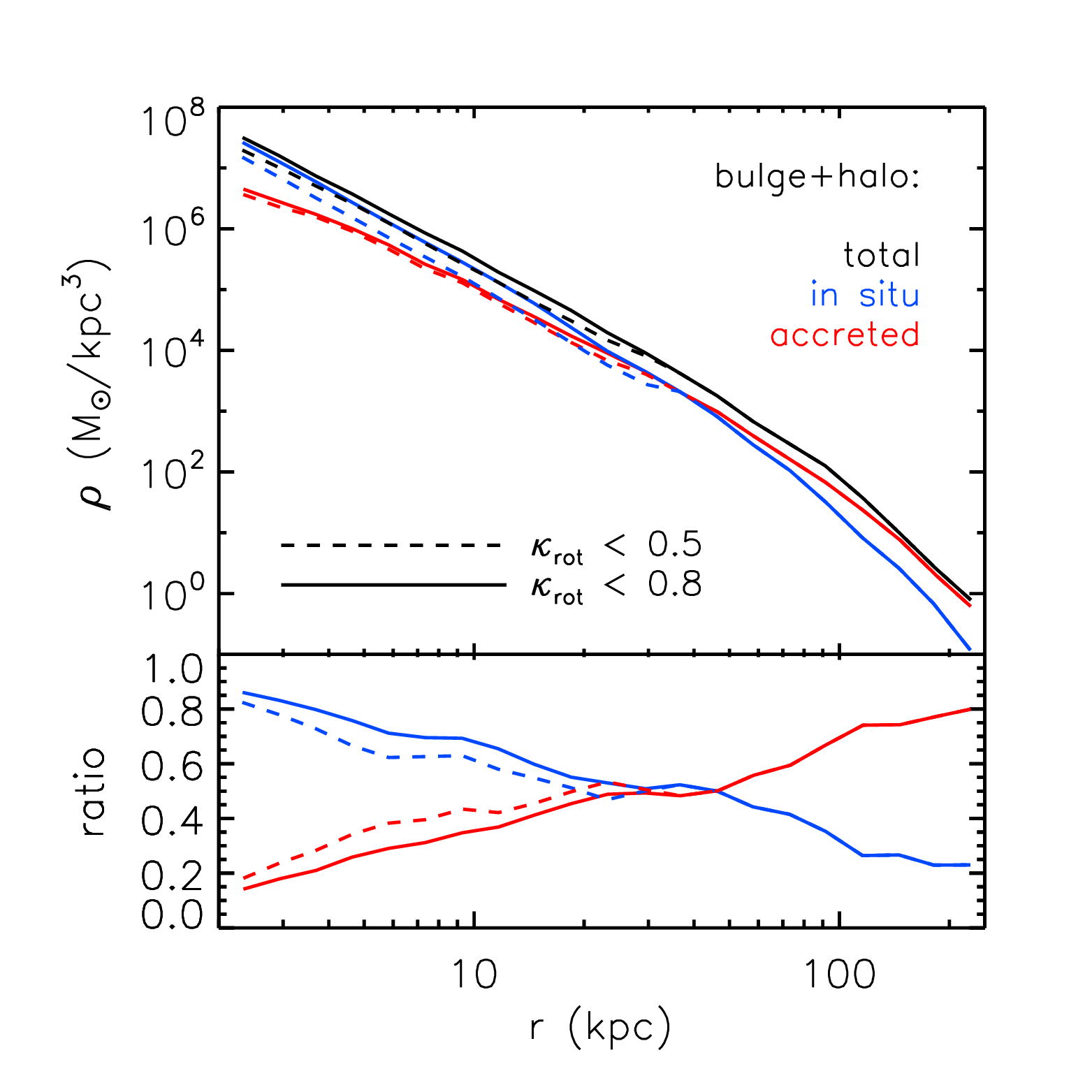}
        \caption{ {\it Left:} The normalised stacked distribution of $\kappa_{\rm rot}$ for all stars (black line) for the star-forming gas (red dashed line) within 30 kpc in the ARTEMIS simulations.  The histograms have be arbitrarily normalised to have a peak value of 1.  {\it Right:} The median, spherically-averaged stellar density profiles of the bulge+halo stellar components: total (black curves), in situ (light blue curves) and accreted (red curves), respectively.  Solid lines correspond to the default criterion for the selection of bulge+halo components, $\kappa_{\rm rot} < 0.8$ (as in right panel of Fig.~\ref{fig:densprof}). Dashed lines correspond to $\kappa_{\rm rot} < 0.5$.  The bottom sub-panel shows the ratios of the in situ and accreted components with respect to the total.}
    \label{fig:densprof_krat}
\end{figure*}

In the right panel of Fig.~\ref{fig:densprof_krat}, we plot the median bulge+halo density profiles for the default $\kappa_{\rm rot}$ < 0.8 (solid lines) and the more conservative $\kappa_{\rm rot}$ < 0.5 (dashed lines) selections.  We also show the (median) separate contributions from the in situ and accreted components (blue and red lines, respectively).  The subpanel shows the median fractional contributions of these components to the total stellar density profile for each of the $\kappa_{\rm rot}$ selections.

As expected, by dropping the $\kappa_{\rm rot}$ threshold from 0.8 to 0.5 more particles are assigned to the disc component and the remaining bulge+halo profile is reduced in the inner regions (compared the dashed black line to the solid black line in the top right panel).  The accreted component of the bulge+halo is mostly unaffected by the change in selection, so the reduction in the total is mostly due to a reduction in the in situ bulge+halo component.  Consequently, the ratio of in situ to accreted contributions changes somewhat (see bottom subpanel) by changing the threshold between bulge+halo and disc.  Overall, though, the effects are fairly modest.  And we view 0.5 as being the lowest value that makes sense for $\kappa_{\rm rot}$, since below this a particle would be supported more by dispersion than ordered rotation.

In the left panel of Fig.~\ref{fig:densprof_krat} we show stacked distributions of $\kappa_{\rm rot}$ for all stellar and star-forming gas particles within 30 kpc for all 42 \textsc{ARTEMIS} haloes.  From visual inspection, the star-forming gas particles are virtually all contained within a thin disc component and, according to the left panel, typically have $\kappa_{\rm rot} > 0.8$.  Thus, as they are born, star particles will have the same $\kappa_{\rm rot}$ distribution, motivating our fiducial threshold.  Of course the star partcles can later be dynamically heated and become part of the bulge+halo component.

\bsp	
\label{lastpage}
\end{document}